\documentclass[12pt, a4paper, onecolumn]{article}
\usepackage[margin=20mm]{geometry}
\usepackage{subfig}
\usepackage{authblk}
\usepackage[dvipdfmx]{graphicx}
\usepackage{wrapfig}
\usepackage{amsmath}
\usepackage{amsthm}
\usepackage{mathtools}
\usepackage{amssymb}
\usepackage{amsfonts}
\usepackage{braket}
\usepackage{mathrsfs}
\usepackage{bm}
\usepackage{here}
\usepackage{bbm}
\usepackage[style=numeric-comp,sorting=none,sortcites=true]{biblatex} 
\usepackage[colorlinks=true]{hyperref}
\usepackage{indentfirst}
\usepackage{booktabs}
\usepackage[noabbrev]{cleveref}
\usepackage[figurename=FIGURE]{caption} 

\usepackage{cleveref}
\crefname{figure}{Figure}{Figures}
\Crefname{figure}{Figure}{Figures}
\bibliography{arXiv}

\begin{document}
\title{Implementation and verification of coherent error suppression using randomized compiling for Grover's algorithm on a trapped-ion device}
\author[1]{Masatoshi Ishii}
\author[2]{Hammam Qassim}
\author[3]{Tomochika Kurita}
\author[2]{Joseph Emerson}
\author[1]{Kazunori Maruyama}
\author[1]{Hirotaka Oshima}
\author[1]{Shintaro Sato}
\affil[1]{Quantum Laboratory, Fujitsu Research, Fujitsu Limited. 10-1 Morinosato-wakamiya, Atsugi, Kanagawa, Japan 243-0197}
\affil[2]{Keysight Technologies Canada, 137 Glasgow St, Kitchener, ON, Canada, N2G 4X8}
\affil[3]{Quantum Laboratory, Fujitsu Research of America, 4655 Great America Pkwy, Suite 410, Santa Clara, CA, United States, 95054}

\maketitle
\begin{abstract}
In near-term quantum computations that do not employ fault tolerant error correction, noise can proliferate rapidly, corrupting the quantum state and making results unreliable. These errors originate from both decoherence and control imprecision and the latter can manifest as coherent error that is especially detrimental. In the pre-fault tolerant setting, previous work has shown that different error suppression methods have shown promising complementary advantages but highly variable performance under different algorithmic and error model conditions.  
Here, we evaluate the effectiveness of several error  suppression methods under  varying algorithmic settings, both theoretically with numerical simulations and experimentally on a trapped-ion quantum computer. For our case study, we explore a range of output states under Grover's algorithm quantum circuits containing up to 10 qubits and 28 two-qubit gates with varying output state features.
We explore the complementary effectiveness of randomized compiling and algorithm error detection, where the latter is realized via post-selection on ancillary qubits that ideally return to the ground state at the end of each circuit. In all settings, combining randomized compiling and error detection yields the largest suppression of error, indicating that these methods are most effective when combined to extend the capabilities of near-term devices for moderately deep circuits.
We demonstrate for the first time significant suppression of coherent error on a trapped-ion platform, and moreover achieve this outcome \textit{via cloud access}.
However our results highlight that the degree of error suppression depends sensitively on the nature of the error model and the algorithm instance. 
\end{abstract}

\def\thefootnote{\fnsymbol{footnote}}
\footnote {This work has been submitted to the IEEE for possible publication. Copyright may be transferred without notice, after which this version may no longer be accessible.}
\section{Introduction}
\label{sec:introduction}
Quantum computers are widely expected to outperform classical computers in certain tasks. To execute useful quantum algorithms reliably at scale, fault-tolerant implementations are required.
However, such implementations remain beyond the reach of current hardware.
In the interim, noisy intermediate-scale quantum (NISQ) devices suffer from significant error rates, making computational outputs increasingly unreliable with increasing circuit depth and qubit count. Nevertheless, recent advances in atomic qubit technologies have led to record qubit lifetimes on the order of seconds and extremely high-fidelity multi-qubit entangling gates, including native gates that act on more than two qubits. These developments raise the possibility of performing classically prohibitive computations using on the order of 100 atomic qubits, with moderately deep circuits, and \textit{without} full-fledged error correction.

In this pre-fault-tolerant regime, and when the circuit depth remains comfortably within the qubit coherence time, decoherence takes a back seat and coherent errors can dominate over stochastic noise. Coherent errors, often arising from limited control precision or crosstalk, can be more damaging than incoherent noise at the same fidelity.
Consequently, mitigating or suppressing coherent errors is critical for obtaining meaningful results from deeper NISQ circuits.

A variety of quantum error suppression and mitigation methods have been proposed to address these challenges, including zero-noise extrapolation \cite{zne_1}, error detection \cite{post_selection, post_processing_protocol}, probabilistic error cancellation \cite{paraerrorcancel_zhang} and randomized compiling \cite{emerson_rc}, each with distinct performance gains, resource costs, and assumptions. Randomized compiling, for example, is known to be particularly effective for suppressing coherent components of the error model \cite{rc_1,rc_pauli_frame, rc_qft_exp} by reducing it to a stochastic form, and the degree of error suppression from randomized compiling depends has been shown to depend not just on the nature of the error model but also on the spread of the output distribution produced by the algorithm \cite{rc_qft_exp}. 
Whereas randomized compiling provides these benefits with negligible overhead, with only a constant number of randomizations, error mitigation methods typically require exponential resource costs in terms of experimental executions \cite{zne_1,PEC}. Randomized compiling is also a valuable preparatory component of many error mitigation schemes because it validates the assumption of stochastic errors that is required for the validity of a linear (rather than quadratic) extrapolation to the zero-error limit \cite{zne_1,PEC, QMLEM}.
The synergetic effects from combining randomized compiling with error mitigation have been recently explored, for example, under purification methods in the quantum imaginary time evolution algorithm \cite{rc_1} and under the zero-noise extrapolation scheme in  the variational quantum eigensolver calculations \cite{kuritaznerc}. A better understanding of the interplay between these error mitigation methods, their dependence on both the nature of the physical error model and on algorithm-specific features, and the resulting degree of error suppression, is primary motivation of the present study.

In this work, we implement and study the synergies of error suppression techniques on a set of instances of Grover's search algorithm implemented on a trapped-ion quantum device. Grover’s algorithm is fundamental algorithm for quantum computing: it provides a quadratic speedup over classical search \cite{grover_algorithm,grover_algorithm2} and is used as a subroutine in numerous quantum algorithms. Additionally the quantum amplitude amplification (QAA) subroutine is a core routine that also appears in many other contexts \cite{qaa_ref0,qaa_ref3,qaa_ref1, qaa_ref2,qaa_ref4}, making our findings broadly applicable across broad algorithm use-cases. Implementations of Grover's algorithm on various quantum devices have been reported \cite{grover_3qubit_exp,grover_exp,grover_ibm_exp} previously, however, the effects of error suppression on this algorithm have not yet been explored and are not well understood.
We investigate randomized compiling alongside error detection to suppress coherent errors. We verify the role of coherent noise through both simulations and experimental data, and explore native-gate transpilation to minimize gate counts. Our experiments on quantum circuits with up to 10 qubits and 28 two-qubit gates show that these two methods work in concert, resulting in a more significant reduction in algorithmic error when used together, but to a degree that depends on the specific instances of the algorithm. Taken together, our results underscore the importance of coherent-error suppression strategies for near-term devices and demonstrate that combining randomized compiling and error detection can significantly enhance the reliability of moderately deep quantum algorithms on trapped-ion platforms. A key finding is that the extent of error suppression depends sensitively on the interplay between the error model and algorithmic instance.
Application-level demonstrations of coherent error suppression by randomized compiling have been reported previously on superconducting devices~\cite{rc_qft_exp, kim_pauli_twirling}.  In this work, we demonstrate for the first time significant suppression of coherent error on a trapped-ion platform, which is significant because trapped-ion platforms can have a completely different error model and, moreover, we achieve this outcome \textit{via cloud access}.

\section{Experimental and simulation procedures}
\label{sec:experimental and simulation procedures}
\subsection{Implementation of Grover's algorithm on a quantum device}
Grover's algorithm comprises three stages: initialization, oracle circuit and QAA circuit.
The initialization circuit prepares the uniform superposition \(\ket{+}^{\otimes n}\).  The oracle is a unitary operation which marks specific bitstrings by imprinting a phase \(\ket{x} \rightarrow -\ket{x}\) for each bitstring \(X\) from a set called the solution set. The QAA circuit amplifies the amplitude of the marked bitstrings. Together, the oracle and QAA circuits form the Grover iteration, which is repeated a number of times to amplify the probability of obtaining a marked bitstring upon measurement.
In quantum circuit implementations of Grover's algorithm, the number of quantum gates required for the oracle circuit varies significantly depending on the marked solution set. For example, implementing a six-qubit Grover's algorithm quantum circuit requires at least one five-controlled-\(Z\) (\(C5Z\)) gate.
The detailed implementation of this circuit on the quantum device is explained in the appendix.

In this work, we study the effect of error suppression on Grover's algorithms using oracles drawn uniformly from an ensemble. We use the total variation distance (TVD) to the ideal circuit output as a measure of performance. The number of two-qubit gates required to implement an oracle greatly affects the overall fidelity, as two-qubit gates are significantly noisy compared to other circuit components. 
An oracle marking a single \(n\)-bit string requires one application of a single-qubit \(Z\) gate controlled on \(n-1\) qubits \cite{grover_3qubit_exp}.
For two or more marked solutions, the number of two-qubit gates varies significantly depending on the structure of the flagged bitstrings, for example, whether they form the marked solution space of a polynomial of a small degree.
This implies that the algorithm encounters different amounts of noise depending on the marked solution set.

\subsection{Quantum error suppression}
In this section, we explain the quantum error suppression methods of randomized compiling and error detection.
\subsubsection{Randomized compiling}
Randomized compiling suppresses coherent noise by randomizing gate implementations while preserving the ideal unitary of a quantum circuit. Random single-qubit gates are inserted into the circuit together with compensating gates later in the sequence, such that their net effect cancels. When averaged over multiple randomized instances, coherent errors are converted into an effective stochastic noise model, thereby reducing their impact on algorithmic performance \cite{emerson_rc}. For an intuitive explanation of the mechanism behind this, see the supplemental material in \cite{rc_qft_exp}.

In practice, randomized compiling is implemented by inserting randomly chosen single-qubit Pauli gates before and then a locally computed correction gate after each two-qubit gate or more generally any `hard' gate in the canonical gate set \cite{emerson_rc}. In this way randomized compiling can be applied to suppress coherent errors in NISQ algorithms and any other universal circuits. Critically, the canonical form and the locally computed correction gate allows randomized compiling to overcome the limitation of Pauli frame randomization \cite{Knill, Kern}, which is restricted to Clifford circuits that allow efficient tracking of the Pauli frame.  Corresponding single-qubit randomizations are also applied to qubits not participating in a given two-qubit gate, in order to maintain consistency across the circuit and mitigate idling and crosstalk effects. Each randomized circuit is logically equivalent to the original circuit but differs in its physical gate realization. In this work, we use 10 randomized circuit instances both in simulations and experiments.

\subsubsection{Error detection}
Error detection is a post-selection technique in which errors are identified by measuring qubits whose ideal final states are known in advance. After measurement, outcomes which deviate from the ideal values indicate the occurrence of errors and are discarded.
In Grover's algorithm, one can use various ancilla-assisted techniques to significantly reduce the depth and gate count by utilizing certain decompositions of the multi-controlled-\(Z\) gates \cite{CnXgate_decomposition_with_ancilla}. In these schemes the ancilla qubits begin and end in the \(\ket{0}\) state -- making error detection especially natural.

While the probability of passing the post-selection filter decreases exponentially with the number of qubits, in our experiment we note a survival ratio ranging between 70\% and 80\%. 
This result, based on the two-qubit gate error rate \(\epsilon\), is mainly due to the active ancilla qubits. In general, if the number of two-qubit gates is \(N_{g}\), the probability of deviating from the ideal state scales roughly as \(P_{pass}\approx(1-\epsilon)^{N_{g}}\).
The device used in the experiment has \(\epsilon\) of 1\%, and calculating this with the QAA gate count of 25 gives \(P_{pass}\approx0.78\). This estimate agrees well with our observed results. For example, if \(\epsilon\) drops to 0.1\%, a comparable post-selection survival rate can be achieved even with 250 gates, allowing error detection to be applied to larger circuits at a practical sampling cost.
It is also possible to use the ancilla-assisted decompositions in a recursive way, and use a strategy for enhancing the probability of passing the post-selection. Another reason this form of error detection works well in tandem with the ancilla-assisted decompositions is that in these decompositions the ancilla qubits are by far the most active ones, participating in every two-qubit gate in \Cref{fig:rptoffoli_decomp_cir,fig:toffoli_decomp_cir,fig:c5z_gate_decomp}. By using them to detect errors we significantly boost the overall fidelity of the entire algorithm.

\subsection{Demonstration of quantum error suppression effects}
Experiments with the six-qubit Grover’s algorithm were performed on the trapped-ion quantum processor IonQ Aria, which provides 25 fully connected qubits \cite{ionq_aria}. At the time of the experiment, the reported average fidelities of single- and two-qubit gates were 0.9998 and 0.99, respectively, with a state-preparation-and-measurement (SPAM) fidelity of 0.9948. The relaxation time \(T_1\) and coherence time \(T_2\) were approximately 100~s and 1~s, respectively, while the gate durations for single- and two-qubit gates were 135~\(\mu\)s and 600~\(\mu\)s. These parameters were also used in the simulation of decoherence, as described later.
The experiments were conducted via Amazon Braket using IonQ’s verbatim compilation mode, which allowed explicit control over qubit allocation and gate implementation. We implemented the six-qubit Grover’s algorithm with 16 marked solutions. For this case, 30 different oracle instances were randomly selected from the 60 oracles that can be implemented using three controlled-\(Z\) (\(CZ\)) gates. A total of 10 qubits were used: six data qubits and four ancilla qubits, the latter used to decompose the \(C5Z\) gate in the QAA circuit.
For randomized compiling, 10 logically equivalent randomized circuit instances were generated. Experiments without randomized compiling used 1000 shots per circuit, while experiments with randomized compiling used 100 shots per randomized instance, resulting in the same total number of shots.
Our paper focuses on relatively small-scale circuits and short circuit depth in order to reduce the impact of decoherence, which on noisy hardware could obscure the more subtle effect of miscalibration. By working in the regime where the overall circuit fits comfortably within the coherence time, we are able to isolate the effect of coherent errors.

\subsection{Simulation of quantum error suppression effects}
As mentioned previously, the number of two-qubit gates required to implement an oracle circuit can vary significantly depending on the structure of the marked  set. In our simulation, this gate-count dependence was removed by replacing the oracle circuit with an ideal unitary matrix. Simulations were additionally performed by calculating Born rule probabilities, which corresponds to an effective infinite number of shots.
In the six-qubit Grover’s algorithm, where there are \(2^6=64\) possible marked states, the QAA circuit produces equivalent behavior for \(r\) and \(64-r\) marked states; consequently, when \(r \geq 32\), the algorithm no longer amplifies the marked set as intended.
For each number of marked solutions up to 31, we simulated 1000 different oracle instances selected uniformly at random. In these simulations, over-rotation noise was used to model coherent errors, while energy relaxation and amplitude damping noise based on the quantum device parameters were included to model decoherence.

\section{Results and discussion}
\label{sec:results and discussion}
To evaluate the effectiveness of quantum error suppression, performance was assessed using the standard metric based on the statistical distance between two probability distributions, TVD \cite{rc_qft_exp}. TVD \(d_{TV}\) is calculated from Eq.~\eqref{eq:tvd}.
\begin{gather}
    d_{TV}({\mathcal{P}},{\mathcal{P}}_{ideal})=\frac{1}{2}\sum_{x\in \{0,1\}^n} |{\mathcal{P}}(x)-{\mathcal{P}}_{ideal}(x)|
    \label{eq:tvd}
\end{gather}
where \(\mathcal{P}_{ideal}(x)\) is the ideal probability of measuring a bitstring \(x\) and \(\mathcal{P}(x)\) is the experimentally measured frequency.
\subsection{Analysis of quantum errors based on the number of marked solutions without any error suppression}
\begin{figure}[h]
    \centering
    \includegraphics[scale=0.4]{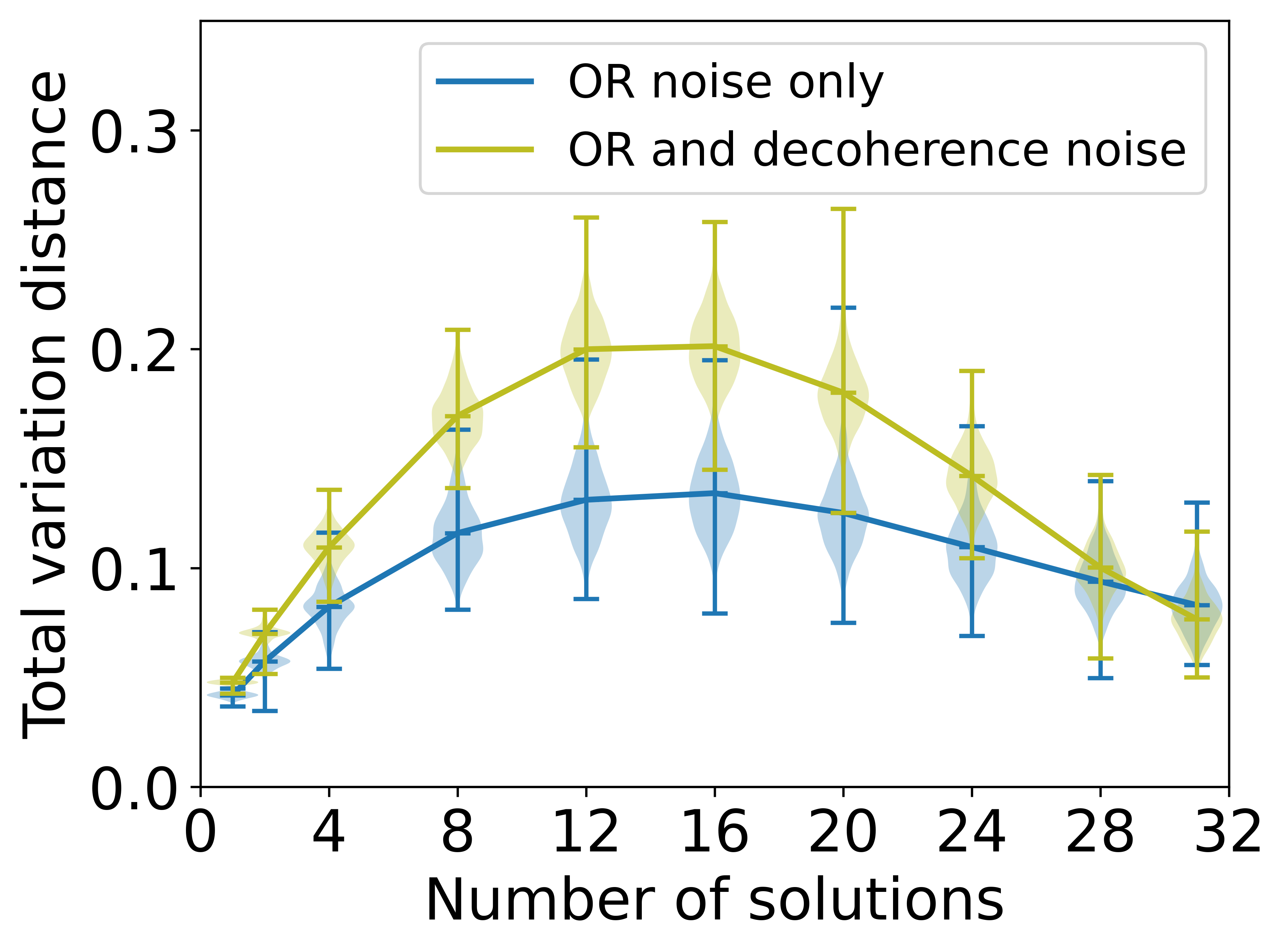}
    \caption{Simulation results of TVD for the number of marked solutions without any error suppression. (The TVD for 1000 randomly selected different marked bitstrings oracle with the same number of marked solutions are plotted. Comparison under two noise models: Over-rotational (OR) noise only, and a combination of OR and decoherence noises.) \textit{Under over-rotation noise, a notable change in TVD is observed, peaking at 16 solutions for both the number of marked solutions and the oracle instances. Adding decoherence noise further increases the TVD values centered around 16 solutions.}}
    \label{fig:NumofSol_tvd_woESonly}
\end{figure}

Figure~\ref{fig:NumofSol_tvd_woESonly} shows a violin plot of the relationship between the number of marked solutions and TVD without any error suppression.
For each number of marked solutions, we calculated 1000 randomly selected different marked bitstring oracles.
In the case where the number of marked solutions is one, there are 64 combinations of marked solutions, so we calculated all 64 oracles.
The simulation results are obtained under two noise models: over-rotation noise only, and over-rotation noise combined with decoherence. The over-rotation noise amplitudes were set to 0.008 for single-qubit gates and 0.08 for two-qubit gates, while decoherence noise parameters were taken from the reported specifications of the IonQ Aria device.
Although not shown explicitly, simulations using between 500 and 2000 randomly selected different oracles yielded consistent results, indicating that the statistical estimates of the mean and spread of the output distributions were stable.

We discuss the following two types of oracle dependencies:\\
1) Dependency based on the number of marked solutions.\\
2) Dependency based on different marked bitstring oracles with the same number of marked solutions.\\
The simulations show that the TVD of Grover’s algorithm in the presence of coherent noise depends strongly on the number of marked solutions.
In the simulation, the oracle circuit is treated as an ideal (noiseless) unitary, so the physical noise sources exist only in the QAA circuit. Even so, oracle dependence  is observed in the final TVD because, when passing through the noisy gates in the QAA circuit, differences in the input states prepared by the oracle change how coherent errors propagate and interfere, which leads to the observed variation in the final TVD.
Details of the implemented circuit are provided in the appendix.

In particular, as shown by the simulation results in Figure~\ref{fig:NumofSol_tvd_woESonly}, for the case of 16 marked solutions, the average TVD increases by approximately a factor of three, with the maximum TVD reaching nearly four times that observed for the single marked solution case under over-rotation noise.
Furthermore, among the 16 marked solutions, the TVD variance exhibits significant dependence on different marked bitstring oracles. The difference between the maximum and minimum TVD values reaches nearly 14 times that observed for a single marked solution under over-rotation noise. Under the combination of over-rotation and decoherence noises, the average TVD increases further, and the difference between the maximum and minimum TVD values slightly increases.
The large spread originates from the coherent nature of over-rotation errors. Unlike stochastic errors whose contributions add incoherently, coherent errors accumulate at the amplitude level and are subsequently amplified or suppressed through interference. Although all oracle instances with 16 marked solutions have the same ideal success probability, they prepare different superposition structures entering the QAA circuit. Consequently, the coherent over-rotation terms propagate differently through the fixed QAA circuit and generate oracle-dependent constructive and destructive interference patterns. This mechanism amplifies the observable TVD variation far beyond the scale observed for a single marked solution.
\begin{figure}
    \centering
    \includegraphics[scale=0.4]{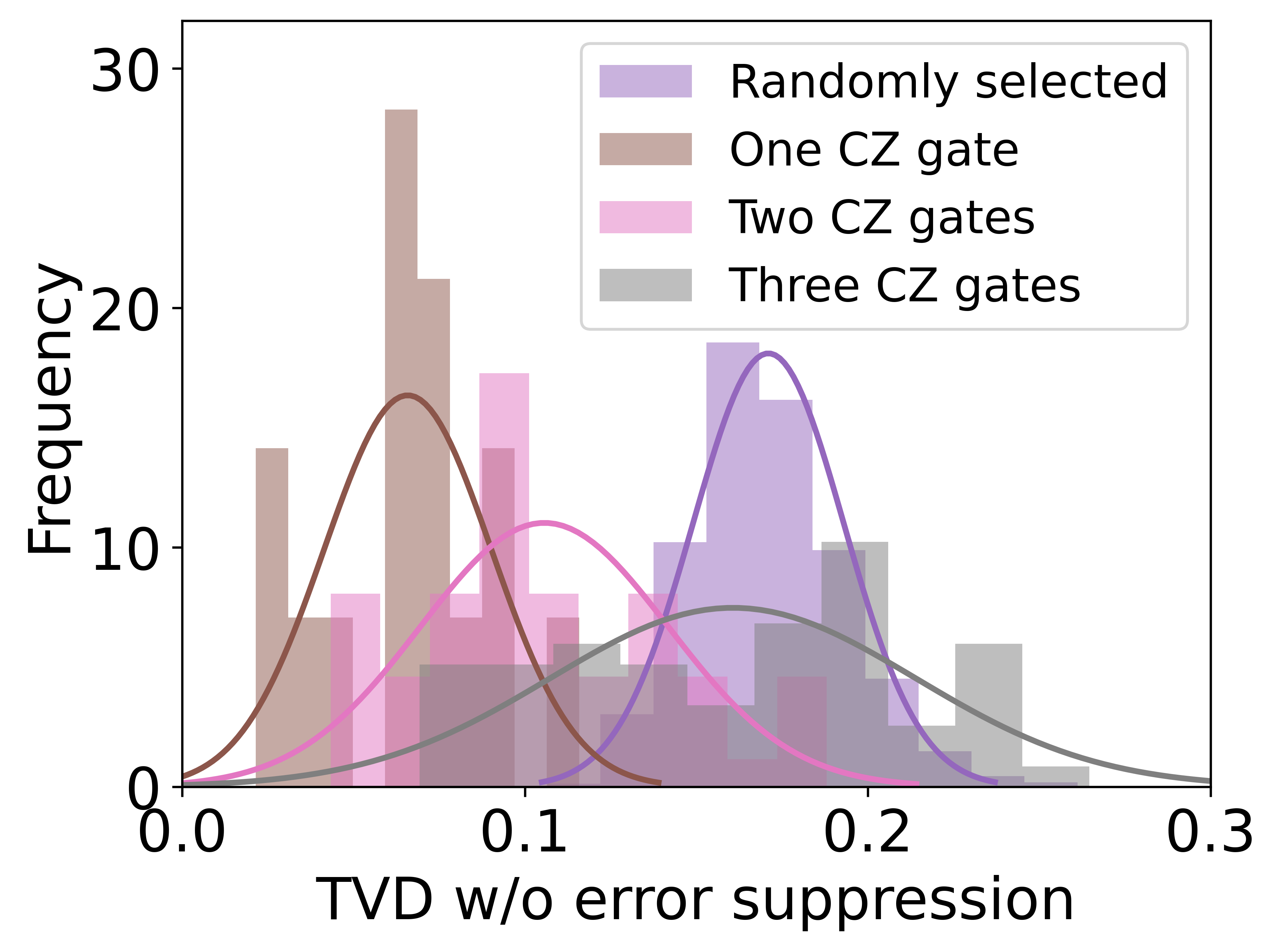}
    \caption{Simulation results on marked bitstring oracle dependency in TVD under only over-rotation noise, with 16 marked solution. (Comparison between a randomly selected marked bitstring oracle and an oracle circuit composed of a different number of \(CZ\) gates.) \textit{Numerical data shows that oracles implementation with three \(CZ\) gates have a large distribution of TVD values and almost the same variance as a randomly selected marked bitstring oracle.}}
    \label{fig:1_3cz_random}
\end{figure}

The oracles with 16 marked solutions have highest sensitivity to coherent errors. Therefore, it is the most compelling case for investigating the quantum error suppression effects of randomized compiling.
To isolate the impact of quantum error suppression we need to compare circuit instances that share the same circuit complexity and depth, so that the cumulative impact of gate error is held constant. Oracle circuits with 16 marked solutions can be constructed with either one, two or three \(CZ\) gates (see the appendix for details), so we next explore the error distributions for each of these three cases.

An example of applying a \(CZ\) gate to qubit i and j is shown in Eq.~\eqref{eq:1_cz}. where \(x = (x_1, x_2, .., x_n)\) varies over all bitstrings. Thus the marked bitstrings are marked solutions to the binary equation \(x_ix_j=1\) \cite{grover_oracle_logic}. Examples of applying two and three \(CZ\) gates are shown in Eq.~\eqref{eq:2_cz} and \eqref{eq:3_cz}, respectively. Therefore, the marked bitstrings are marked solutions to the binary equations \(x_ix_j+x_ix_k=1\) and \(x_ix_j+x_ix_k+x_ix_l=1\), respectively. Where, \(i\), \(j\), \(k\), and \(l\) denote distinct qubit numbers.
\begin{align}
    \sum_x{\ket{x}}  \rightarrow &\sum_x{(-1)^{x_ix_j}\ket{x}} \text{ : One $CZ$} \label{eq:1_cz} \\
    \sum_x{\ket{x}}  \rightarrow &\sum_x{(-1)^{x_ix_j+x_ix_k}\ket{x}} \text{ : Two $CZ$} \label{eq:2_cz} \\
    \sum_x{\ket{x}}  \rightarrow &\sum_x{(-1)^{x_ix_j+x_ix_k+x_ix_l} \ket{x}} \text{ : Three $CZ$} \label{eq:3_cz}
\end{align}

We evaluated the TVD's marked bitstring oracle dependency due to over-rotation noise for quantum circuits that implement six-qubit Grover's algorithm with 16 marked solutions, using oracle decompositions using one to three \(CZ\) gates. The results are shown in Figure~\ref{fig:1_3cz_random}.
The calculations were performed under the same conditions as in Figure~\ref{fig:NumofSol_tvd_woESonly} for the case of over-rotation only.

Oracles constructed using one or two \(CZ\) gates exhibits TVD values confined to a narrower range than the overall distribution obtained from a randomly selected oracle.
However, the TVD distribution obtained from all three-\(CZ\)-gate oracles closely reproduces the observed mean and variance obtained from 1000 uniformly random oracles.
It was shown that increasing the number of \(CZ\) gates with maintaining the number of marked solutions allows for a greater number of different combinations of marked bitstring oracles. 
Therefore, to verify a typical randomly selected marked bitstring oracle, implementing oracle circuits using three \(CZ\) gates is appropriate.
In the quantum device experiment, 30 types were randomly selected and executed from among 60 types of marked bitstring oracles implementable using three \(CZ\) gates.

\subsection{Experimental results of quantum error suppression effects}
\begin{figure*}[h]
    \centering
    \subfloat[Comparison of distributions of TVD for cases without error suppression (ES), with randomized compiling (RC), with error detection (ED), and with both RC and ED. \textit{The experimental results indicate that the error suppression effect increases in the order of RC, ED, and  both RC and ED.}]{\includegraphics[scale=0.4]{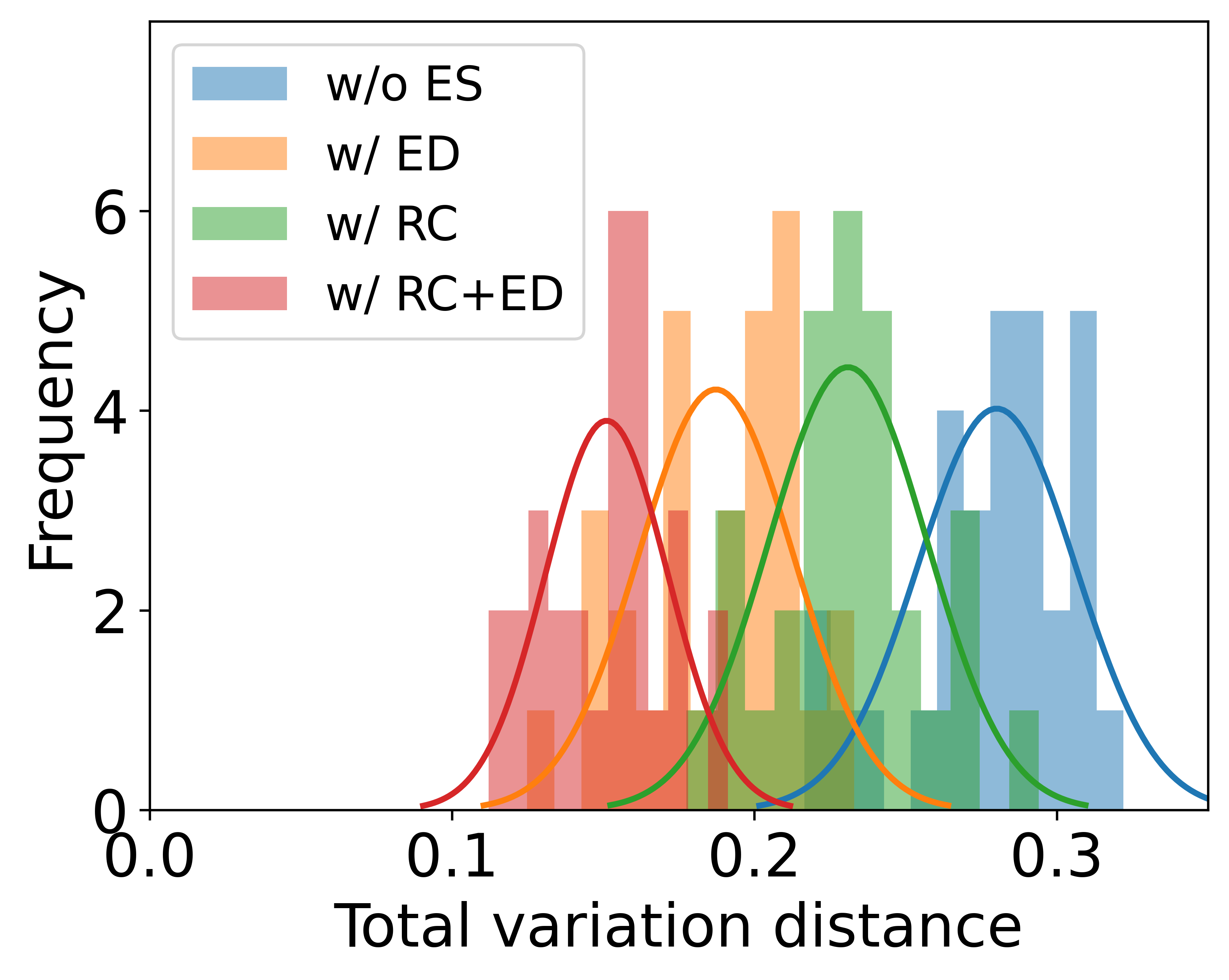}
    \label{fig:ionq_results}}
    \hspace{10mm}
    \subfloat[Comparison of distributions of TVD improvement factors for cases without ES, with RC, with ED, and with both RC and ED.
     \textit{The average improvement factor increases to approximately 1.9 with both RC and ED, indicating that the combined approach provides the largest overall reduction in algorithmic error.}
    ]{\includegraphics[scale=0.4]{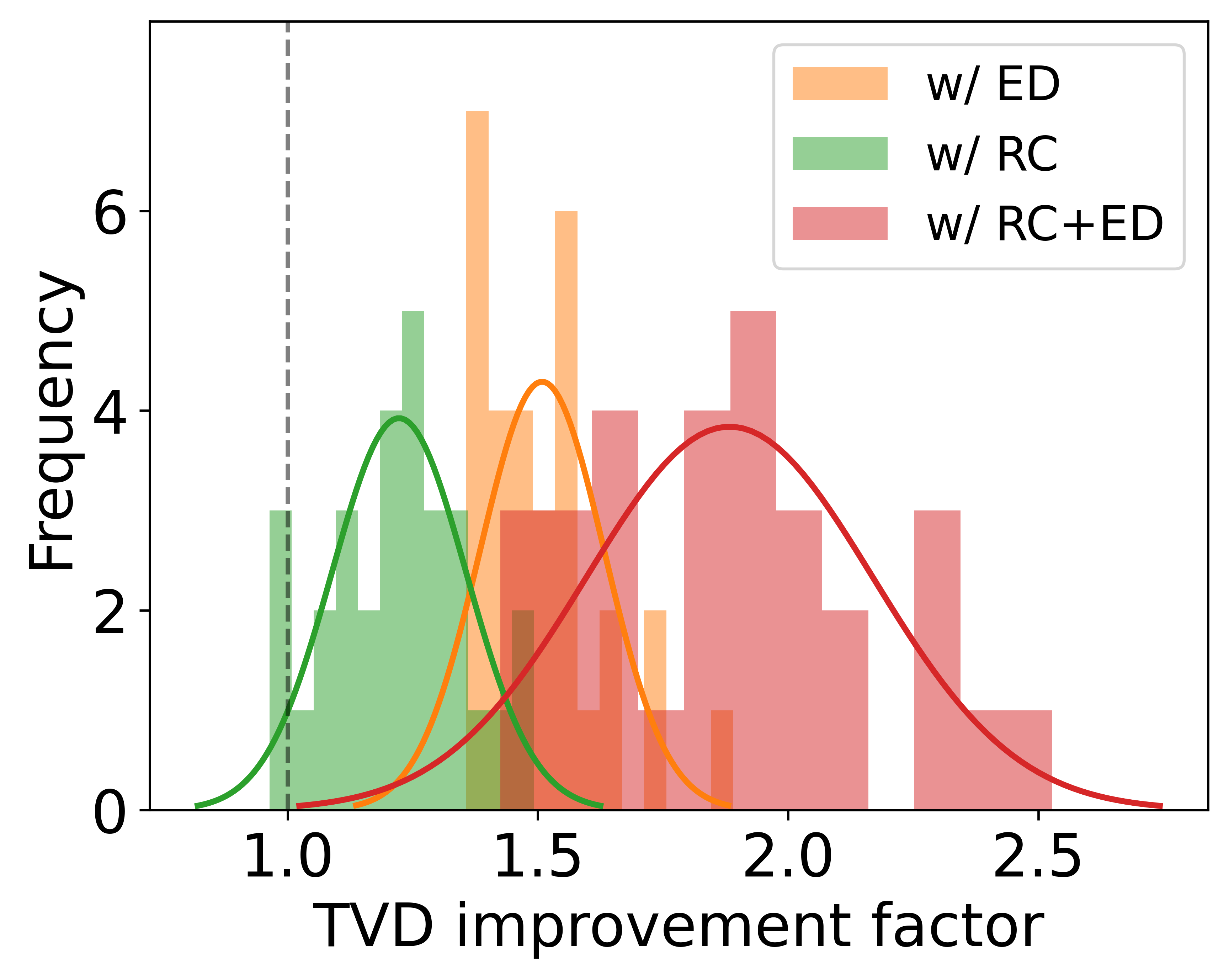}
    \label{fig:ionq_results_impfact}}
    \caption{Experimental results for a circuit implementing three-\(CZ\) gate oracles circuits using a trapped-ion device, with 16 marked solutions and randomly selected different marked bitstring oracles.}
    \label{fig:ionq_results_all}
\end{figure*}

The experimental results of the six-qubit Grover’s algorithm with 10 qubits in total and with 16 marked solutions executed on IonQ Aria are shown in Figure~\ref{fig:ionq_results_all}. The figure presents the TVD values obtained for 30 different marked bitstring oracle instances without error suppression (blue), with randomized compiling only (green), with error detection only (orange), and with the combined application of randomized compiling and error detection (red). As shown in Figure~\ref{fig:ionq_results}, the TVD without error suppression exhibits a strong dependence on marked bitstring oracle, ranging from approximately 0.22 to 0.32.
Because all circuits were implemented using the same numbers of single- and two-qubit gates, this oracle dependence cannot be attributed to differences in gate count.
Applying randomized compiling and error detection leads to a reduction in the typical TVD values, with the lowest TVD generally observed when both techniques are applied together.

Figure~\ref{fig:ionq_results_impfact} summarizes these results using the TVD improvement factor, defined as the ratio of the TVD without error suppression to the TVD obtained after applying error suppression. 
The improvement factor for randomized compiling alone is slightly below 1 for some oracle instances, but its average is approximately 1.2. 
The average improvement factor increases to approximately 1.5 when error detection is applied alone, and to approximately 1.9 when randomized compiling and error detection are combined, indicating that the combined approach provides the largest overall reduction in algorithmic error.

\subsection{Simulation analysis of quantum error suppression effect}
\subsubsection{Analysis of quantum errors dependent on the number of marked solutions with quantum error suppression.}
\begin{figure*}[h]
    \centering
    \subfloat[Under over-rotation noise only.
    \textit{The simulation results show that applying randomized compiling leads to a reduction in both the mean and variance of TVD.}]{\includegraphics[scale=0.4]{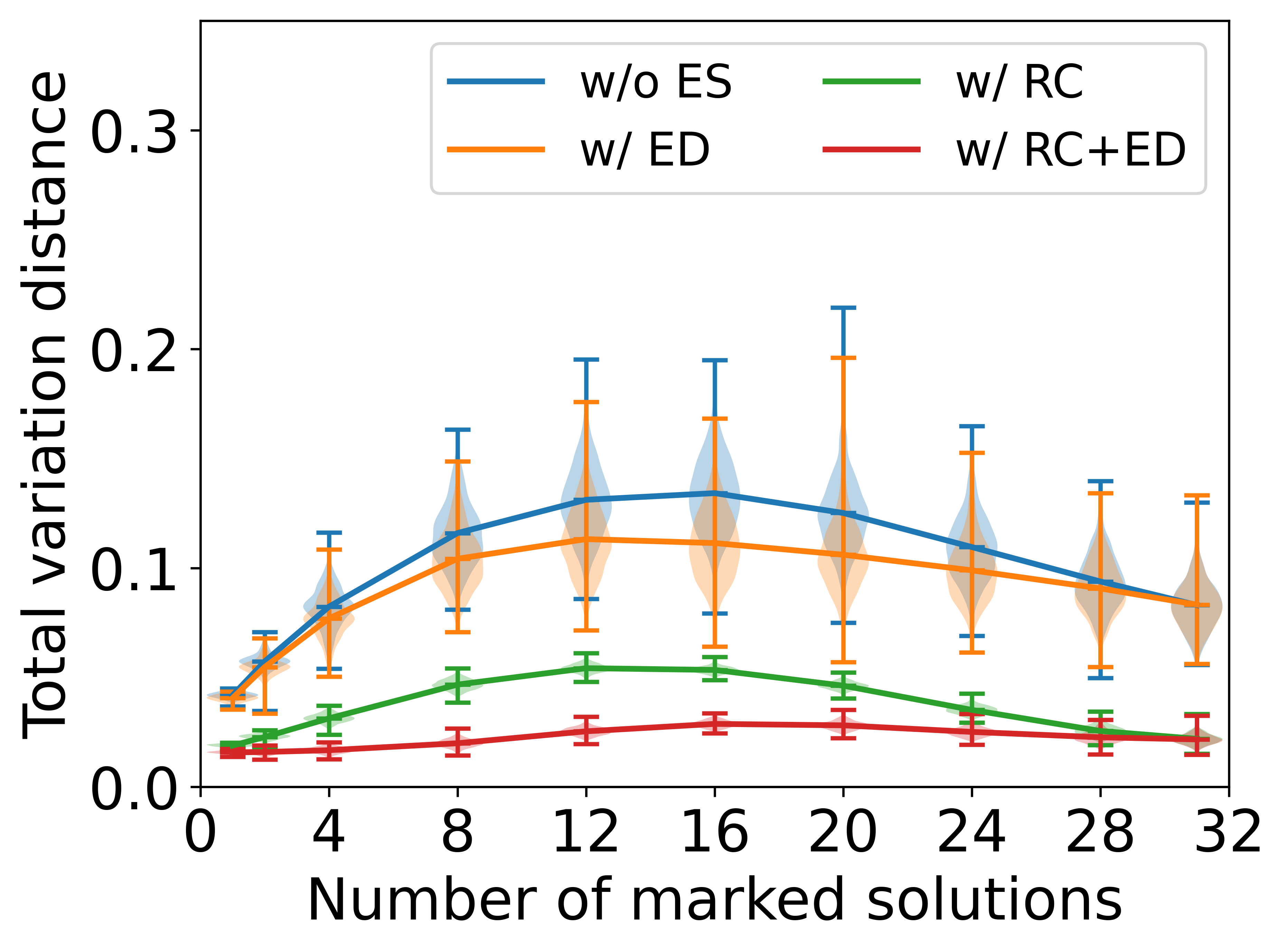}
    \label{fig:NumofSol_tvd1}}
    \hspace{10mm}
    \subfloat[Under over-rotation and decoherence noise.
    \textit{Simulation results show that applying both of randomized compiling and error detection reduces both the mean and variance of TVD even under both over-rotation and decoherence noise.}
    ]{\includegraphics[scale=0.4]{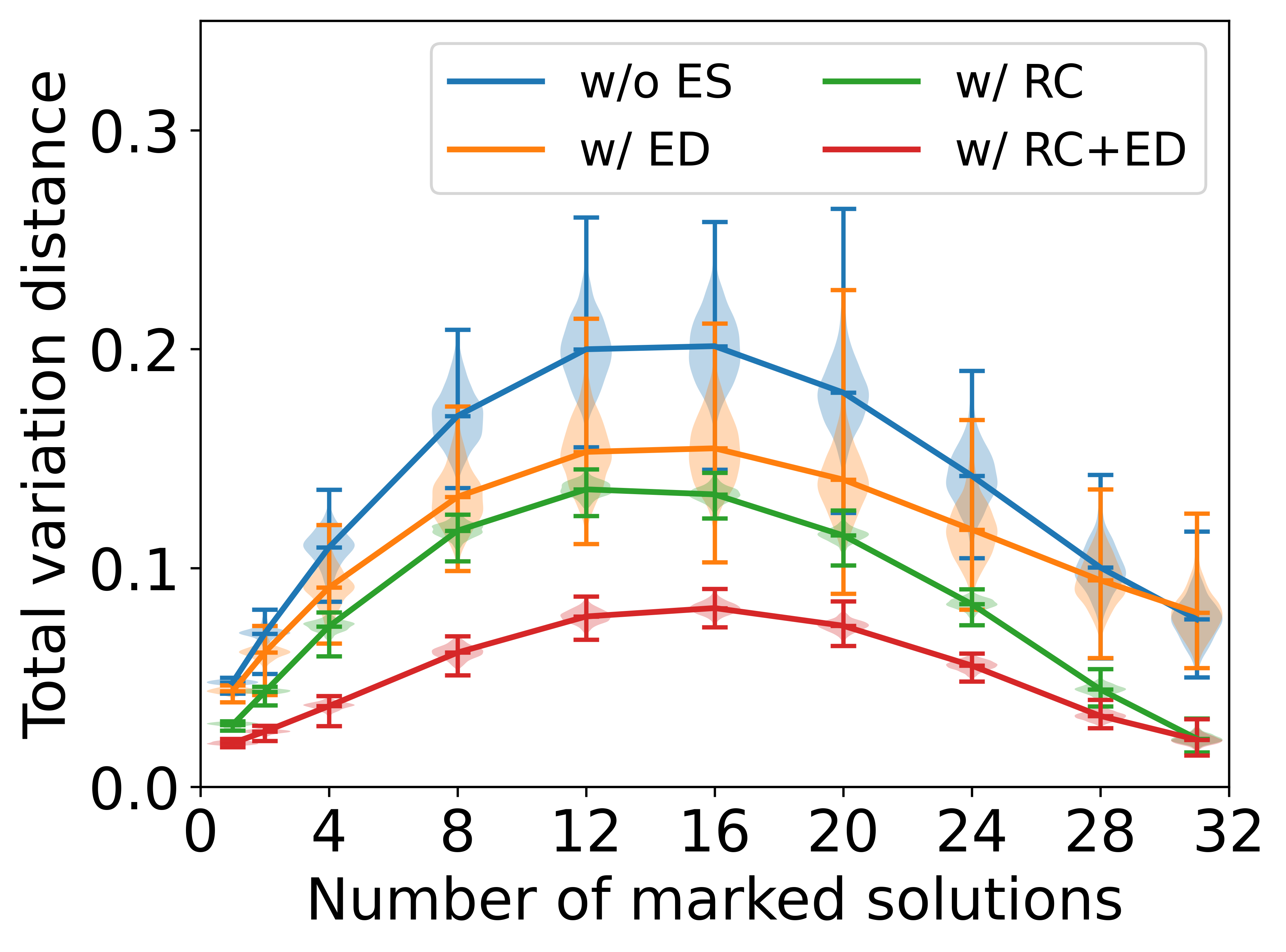}
    \label{fig:NumofSol_tvd2}}
    \caption{Simulation results for TVD variations based on the number of marked solutions for randomly selected marked bitstring oracles for cases without error suppression (ES), with randomized compiling (RC), with error detection (ED), and with both RC and ED. }
    \label{fig:NumofSol_tvd}
\end{figure*}

Figure~\ref{fig:NumofSol_tvd} shows simulation results for TVD variations based on the number of marked solutions for randomly selected marked bitstring oracles for cases without error suppression, with randomized compiling, with error detection, and with both randomized compiling and error detection. 
It is observed that when coherent noise is present, both the mean and variance of TVD strongly depend on the number of marked solutions and the marked bitstring oracle, but the application of randomized compiling leads to a reduction in both the mean and variance of TVD, as shown in Figure~\ref{fig:NumofSol_tvd1}.
From Figure~\ref{fig:NumofSol_tvd2}, when both over-rotation and decoherence noise are included, the overall TVD improvement achieved by randomized compiling is smaller than in the case of over-rotation noise alone.
Nevertheless, randomized compiling still reduces the marked bitstring oracle dependent variance of the TVD under the combined noise model. In contrast, when only over-rotation noise is present, the improvement obtained using error detection alone is smaller than that obtained using randomized compiling.

\subsubsection{Analysis of quantum errors dependent on marked bitstring oracle with quantum error suppression}
The violin plot in Figure~\ref{fig:NumofSol_tvd} provides a compact visualization of how the mean TVD and its marked bitstring oracle dependent variations change as a function of the number of marked solutions. However, this representation does not directly reveal how the TVD associated with a given marked bitstring oracle is modified by the application of error suppression methods.

\begin{figure*}[h]
    \centering
    \subfloat[Under over-rotation noise only.
    \textit{Applying error detection only reduces TVD to a constant value compared to the case without error suppression. In contrast, applying randomized compiling significantly reduces the variation in TVD values dependent on the oracle.}
    ]{\includegraphics[scale=0.4]{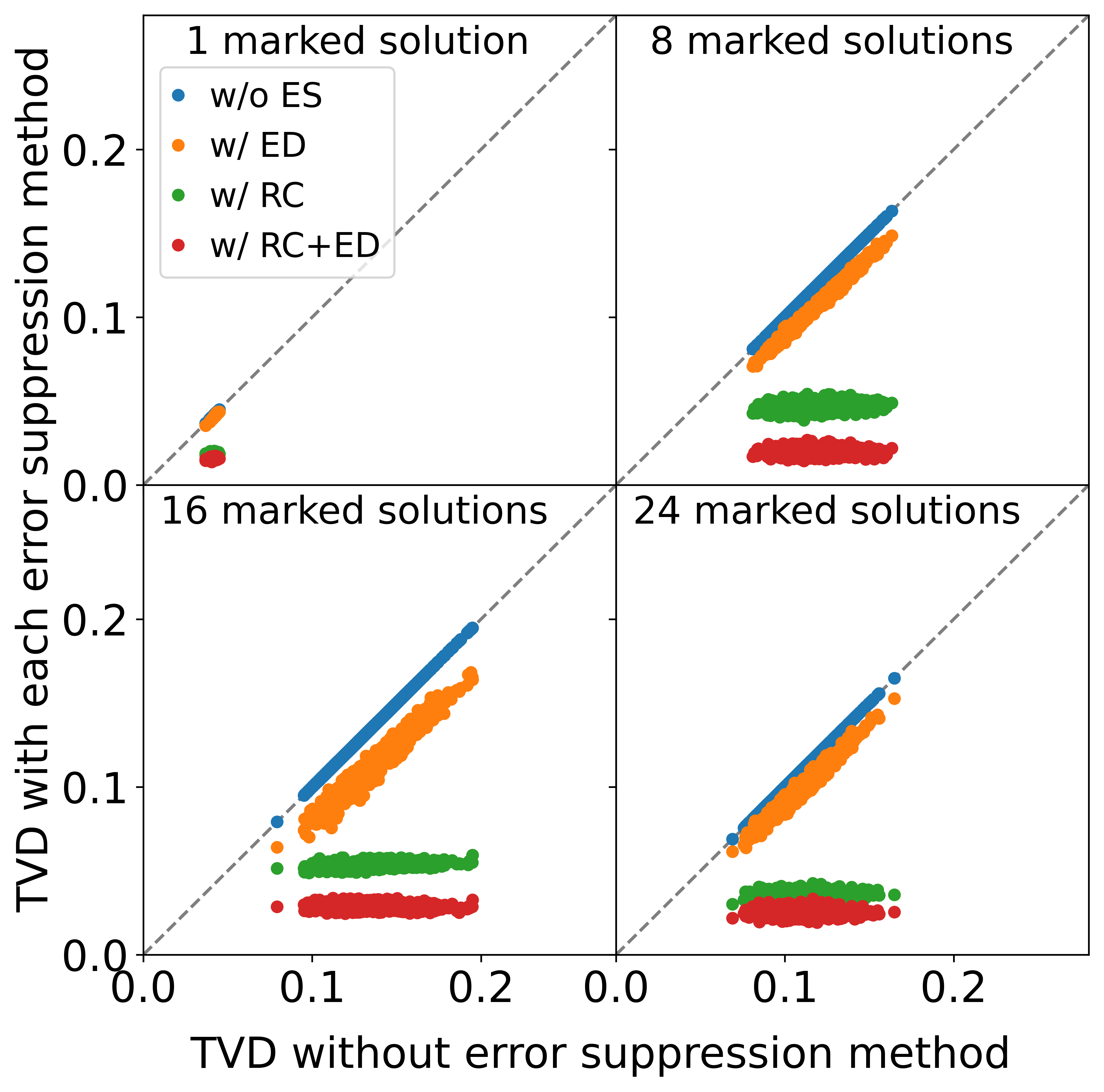}
    \label{fig:es_ana1}}
    \hspace{10mm}
    \subfloat[Under over-rotation and decoherence noise.
    \textit{Under conditions of over-rotation and decoherence noise, the plotted points shift overall to the upper right compared to the case of over-rotation noise alone. This indicates that a similar effect can be expected regarding the impact of each error suppression method on TVD.}
    ]{\includegraphics[scale=0.4]{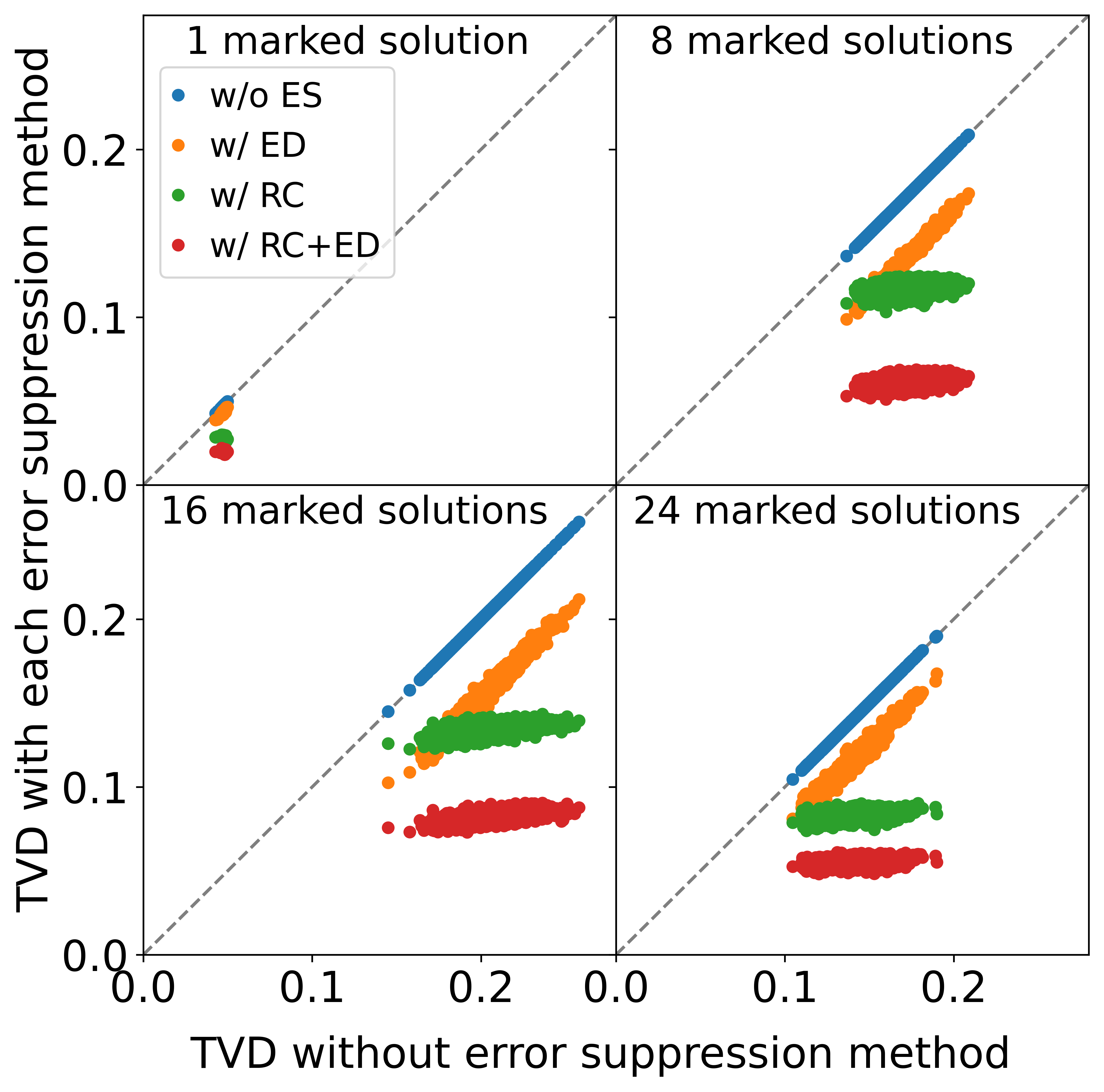}
    \label{fig:es_ana2}}
    \caption{Analysis results of quantum error suppression mechanisms in simulation. \textit{The plot in Figure~\ref{fig:es_ana2} (over-rotation and decoherence) is shifted toward the upper right compared with Figure~\ref{fig:es_ana1} (over-rotation only), indicating that the presence of decoherence increases the baseline TVD in all cases. However, the relative reduction in TVD variance due to RC and the reduction in mean TVD due to ED are consistent with the over-rotation-only case.}}
    \label{fig:es_ana}
\end{figure*}

Figure~\ref{fig:es_ana} presents an alternative analysis highlighting each error suppression effects at each level of the marked bitstring oracle for the number of marked solutions being 1, 8, 16, and 24.
This plot shows the relationship between the TVD value without error suppression (horizontal axis) and the TVD value with and without error suppression (vertical axis).
The blue points plot the TVD for randomly selected each different marked bitstring oracle without error suppression. The dashed line represents the state where the TVD on the horizontal and vertical axes are equal (no improvement), and the blue points are plotted on this line.
The orange, green, and red points show the respective TVD with error detection, with randomized compiling, and with both randomized compiling and error detection, respectively. 
These plots are plotted as upward or downward changes from the TVD value without error suppression for each marked bitstring oracle. Therefore, data points below the blue points indicate a reduction in the TVD value due to error suppression.

As shown by the orange points in Figure~\ref{fig:es_ana1}, applying error detection reduces the TVD from those without error suppression to a constant value, indicating no suppression effect on the variation of TVD values across each marked bitstring oracle.
In contrast, when randomized compiling is applied, as shown by the green points, the marked bitstring oracle dependent spread of TVD values is strongly reduced, leading to a clustering of points with similar TVD values. When randomized compiling and error detection are applied together, the reduction in TVD is further enhanced.

As shown in \ref{fig:es_ana2}, when over-rotation and decoherence noises are both applied, the data points without error suppression, as shown by blue point, are shifted towards the upper right compared to the case under only over-rotation noise.
Similarly, when randomized compiling, as shown by green points, is employed, an upward shift to the right was observed. This behavior suggests that randomized compiling is primarily effective in suppressing over-rotation noise. Meanwhile, when error detection (orange points) was used, the decrease in TVD was enhanced compared with that in the case under over-rotation noise only.
This behavior indicated that error detection exerted an error suppression effect mainly on stochastic noise, such as decoherence noise.
Furthermore, the reduction in TVD achieved by applying both randomized compiling and error detection (red points) compared to randomized compiling alone (green points) was greater than the reduction achieved by applying error detection alone (orange points) compared to no error suppression.

These behaviors are consistent with the interpretation that randomized compiling converts coherent error contributions into an effectively stochastic noise component, which can then be more efficiently suppressed through error detection. As a result, the combined application of randomized compiling and error detection leads to a synergistic improvement in algorithmic performance.

\subsubsection{Comparison of experimental and simulation results}
\begin{figure*}[h]
    \centering
        \subfloat[Experimental results.
        \textit{Re-plotting the experimental data confirmed that similar analysis results to those in Figure~\ref{fig:es_ana} are obtained.}
        ]{\includegraphics[scale=0.4]{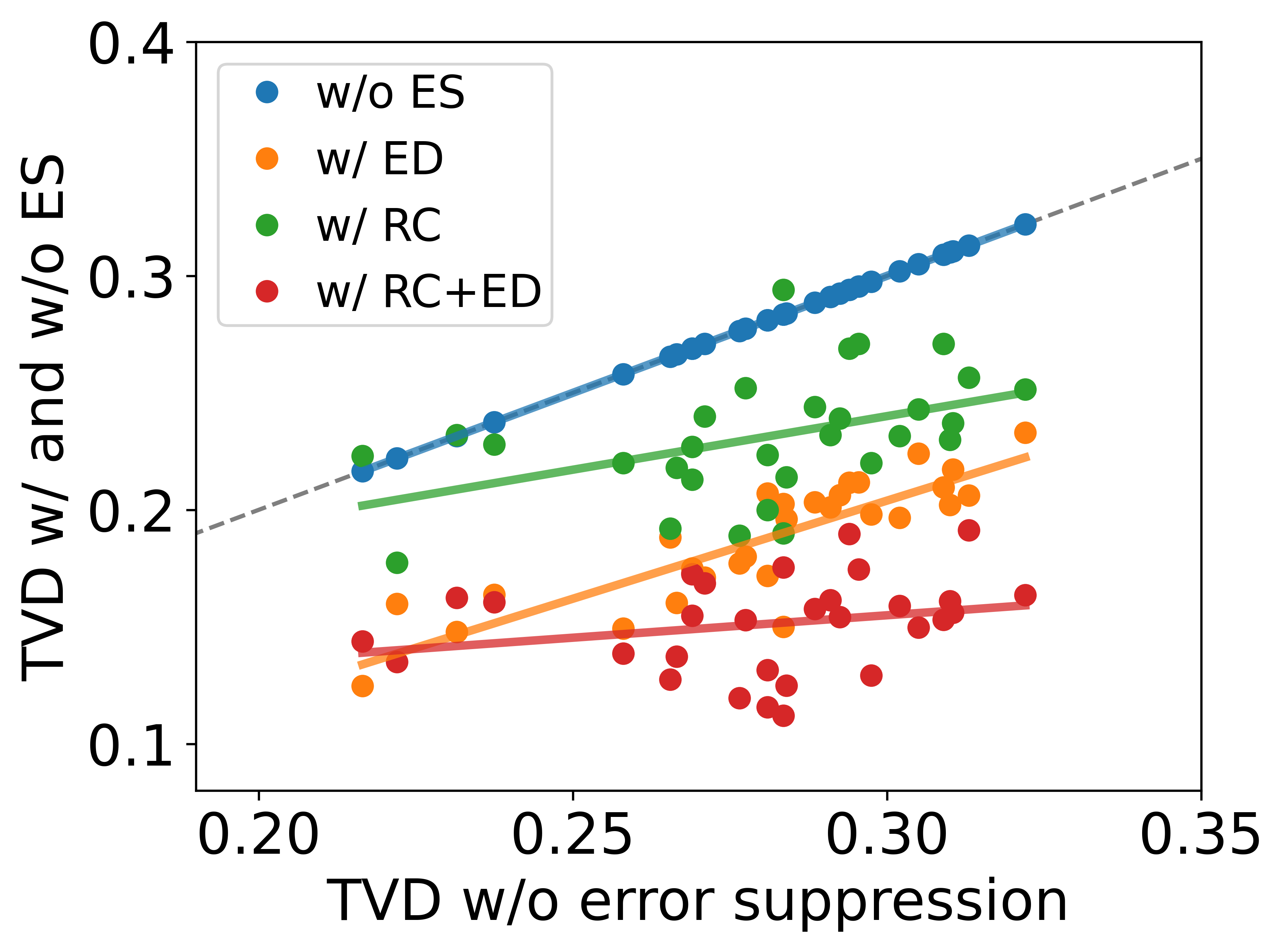}
        \label{fig:es_ana_ionq}}
  \hspace{10mm}
  \subfloat[Simulation results.
  \textit{Both experimental and simulation results show close agreement, demonstrating that randomized compiling can suppress TVD variation caused by marked bitstring oracle dependencies due to coherent noise.}
  ]
  {\includegraphics[scale=0.4]{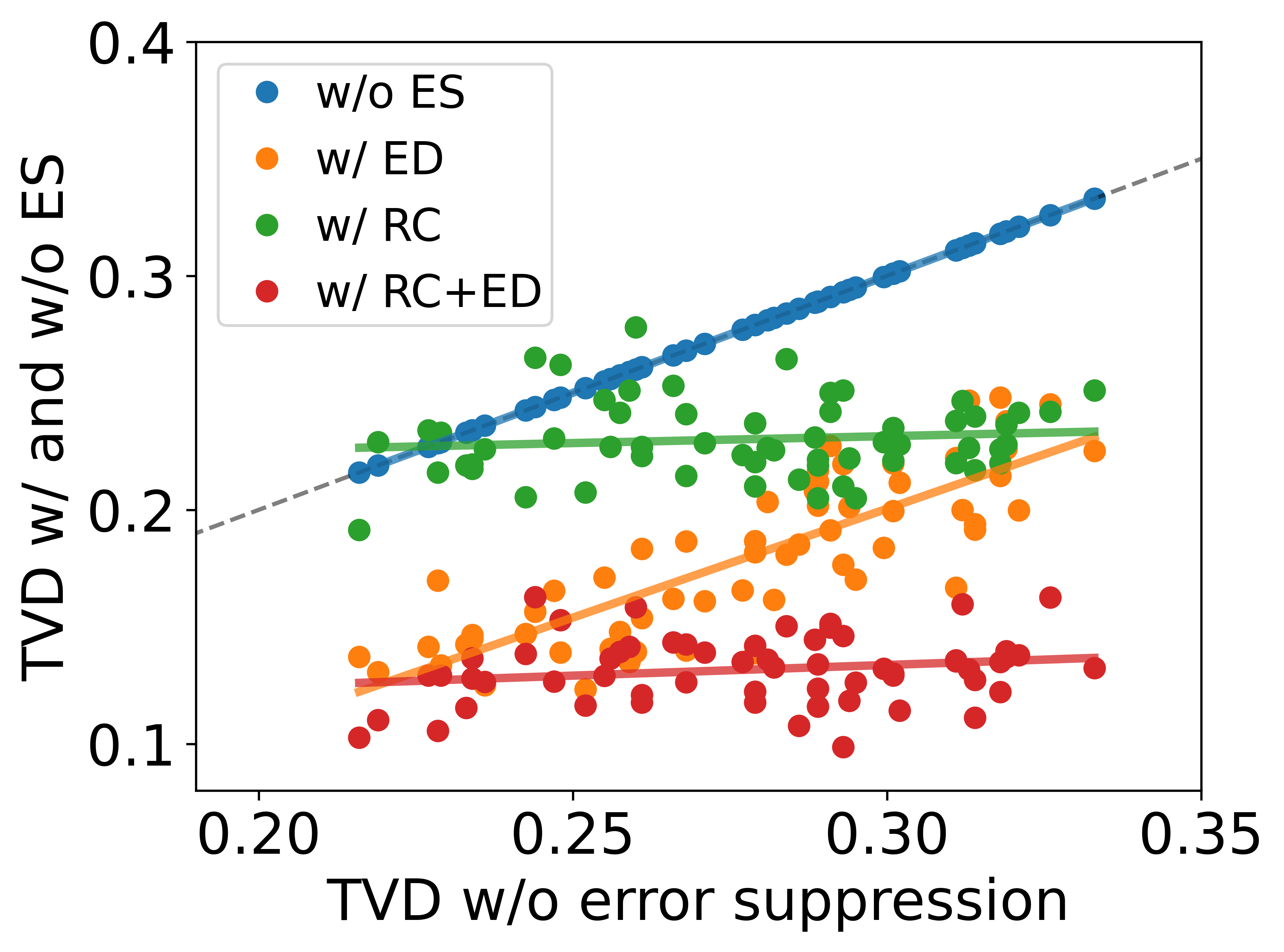}
  \label{fig:es_ana_sim}}
  \caption{Comparison of the error suppression effect between experimental results in a trapped-ion device and simulation results. }
  \label{fig:exp_and_es_ana_sim}
\end{figure*}

Figure~\ref{fig:exp_and_es_ana_sim} shows a comparison between the experimental results obtained on IonQ Aria and the corresponding simulation results. The data shown in Figure~\ref{fig:es_ana_ionq} are identical to those in Figure~\ref{fig:ionq_results}. While earlier simulations used an effectively infinite number of shots to eliminate finite-sampling effects, Figure~\ref{fig:es_ana_sim} presents simulation results obtained using 1000 shots to enable a fair comparison with experiment. Similarly, randomized compiling circuits are simulated using 100 shots for each of the 10 randomizations, matching the experimental protocol.
The simulations included both over-rotation and decoherence noise. Over-rotation noise was applied with magnitudes of 0.008 for single-qubit gates and 0.08 for two-qubit gates. Decoherence was modeled using energy relaxation and dephasing channels, with \(T_1\) and \(T_2\) values reported for the IonQ Aria device. In addition, stochastic Pauli noise with magnitude \(7 \times 10^{-3}\) was applied to the two-qubit gates only. The simulations also incorporated a native gate decomposition consistent with that implemented on the trapped-ion device.
The simulation results exhibit trends closely matching those observed experimentally.
Overall, the close agreement between experiment and simulation supports the validity of the simulation analysis.

\section{Conclusion}
\label{sec:conclusion}

In this work, we systematically investigated the role of quantum error suppression techniques through a both numerical simulation and the experimental implementation of the six-qubit Grover’s algorithm on a trapped-ion quantum device.
We first demonstrated that, in a trapped-ion device, the impact of coherent errors can play a significant role in the implementation accuracy of Grover’s algorithm in a way that depends on the oracle even for circuits that have the same number of two-qubit gates.
These results were demonstrated in both the mean value and variation of the TVD between the ideal and noisy solutions. Numerical simulation results further show that this oracle dependence is correlated with the number of marked solutions and arises primarily from the amplitude amplification circuit. It should be noted that Grover’s search algorithm will be typically applied in settings where the number of marked solutions is not known \emph{a priori}.

We then show that randomized compiling has a significant error-suppression effect on this sensitivity to coherent noise, stabilizing algorithmic performance across different oracle instances. We first shows this numerically and then also show experimentally that this suppression occurs also on the trapped-ion platform. Because randomized compiling can not suppress the incoherent part of the noise, these results imply a significant part of the  error is coherent on the trapped-ion platform.
This appears to be the first experimental demonstration of the synergistic effect of randomized compiling and error detection on a trapped-ion platform, specifically applied to an algorithmic task (Grover’s algorithm).

We further show that randomized compiling, when combined with error detection, enjoys an additional synergistic improvement, yielding the largest overall reduction in algorithmic error, although the degree of error suppression depends sensitively on the number of marked solutions. Our findings illustrate the complementary roles of randomization for coherent-error suppression and post-selection-based error reduction as a function of the oracle, and in some instance that post-selection methods exhibit a strong dependence on the stochastic nature of the error (as guaranteed through randomized compiling). Because error mitigation, including specific error detection strategy employed here, generally requires an exponentially increasing experimental  resource cost (with system size), whereas randomizing compiling incurs only a constant overhead cost, we demonstrate the relative gains from each error suppression strategy (and their combined effectiveness). These results should help inform resource-estimation and performance tradeoffs when planning  the design and implementation of error suppression and mitigation strategies, in general, and in particular for the development of application-oriented benchmarks of hardware performance.

Collectively, our results demonstrate that error suppression tools are not a “one size fits all” solution with uniform error suppression properties, but there is a complex interplay between the nature of the error model and the details of the algorithm instance. These results highlight  the relative importance of constant overhead vs exponential overhead error reduction strategies. 

\section{Appendix}
\label{sec:appendix}
\subsection{Implementation of six-qubit Grover's algorithm on a quantum device}
\begin{figure}[h]
    \centering
    \includegraphics[scale=0.8]{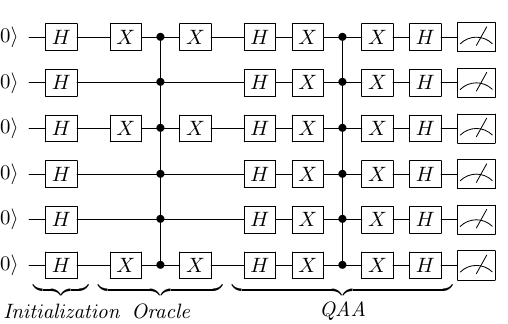}
    \caption{Quantum circuit of six-qubit Grover’s algorithm with one marked solution.}
    \label{fig:grover_1sol_cir}
\end{figure}

The quantum circuit of the six-qubit Grover's algorithm with one solution is shown in Figure~\ref{fig:grover_1sol_cir}. In the initialization circuit, the Hadamard (\(H\)) gate is applied to all qubits to prepare the uniform superposition. The oracle circuit uses \(C5Z\) gates and \(X\) gates to mark specific bitstring with phase. The presence or absence of the \(X\) gate depends on the oracle. A QAA circuit comprises a \(C5Z\), \(H\), and \(X\) gates. Therefore, the Grover's algorithm using one solution requires two \(C5Z\) gates.

\begin{figure}
    \centering
    \includegraphics[scale=0.8]{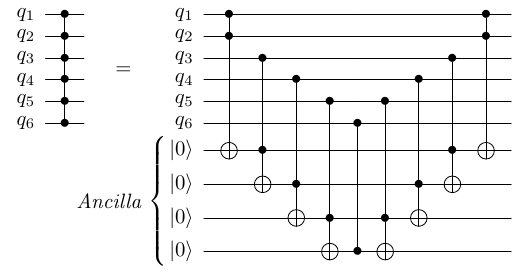}
    \caption{Decomposed quantum circuit of \(C5Z\) gate with eight Toffoli gates and four ancilla qubits.}
    \label{fig:c5z_gate_decomp}
\end{figure}

\begin{figure}
    \centering
    \subfloat[Toffoli gate.]{\includegraphics[scale=0.8]{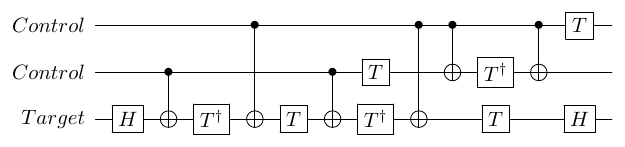}\label{fig:toffoli_decomp_cir}}
    \vspace{0mm}
    \subfloat[Relative-phase Toffoli gate.]{\includegraphics[scale=0.8]{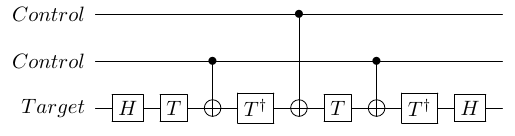}\label{fig:rptoffoli_decomp_cir}}
    \caption{Decomposed quantum circuits of typical Toffoli and relative-phase Toffoli gate.}
\end{figure}

As mentioned above, gate operation on NISQ devices is limited to the single- or two-qubit gates, which requires circuit decomposition. On NISQ devices, the infidelity of a two-qubit gate is about an order of magnitude greater than that of a single-qubit gate. Therefore, the number of the two-qubit gate significantly affects noise generated by the algorithm when implementing quantum circuits in a device.
There are various decomposition techniques for the multi-controlled-\(Z\) gate \cite{CnXgate_decomposition_with_ancilla, CnZgate_decomposition_without_ancilla}. Herein, we use a decomposition technique that employs ancilla qubits and Toffoli gates, as shown in Figure~\ref{fig:c5z_gate_decomp}. The \(C5Z\) gate can be decomposed with four ancilla qubits and eight Toffoli gates along with a \(CZ\) gate. The implementation of a typical Toffoli gate requires six controlled-\(X\) (\(CX\)) gates, as shown in Figure~\ref{fig:toffoli_decomp_cir} \cite{toffoli_to_cx_decomp}. The same logical gate can be realized by replacing the Toffoli gate with a relative-phase Toffoli gate in the decomposition of the multi-controlled-\(Z\) gate \cite{cnz2rptoffoli_gate}.
Using the relative-phase Toffoli gates, the number of \(CX\) gates required for decomposition is reduced from six to three per Toffoli gate, as shown in Figure~\ref{fig:rptoffoli_decomp_cir}. As a result, the \(C5Z\) gate is decomposed using four ancilla  qubits and 25 two-qubit gates (24 \(CX\) and 1 \(CZ\) gates). The \(C5Z\) gate is also required for the QAA circuit, and implementing the Grover's algorithm with one solution on NISQ devices requires 50 two-qubit gates. An ancilla qubit used to decompose the multi-controlled-\(Z\) gate is initially in the \(\ket{0}\) state and will return to the \(\ket{0}\) state after execution if there is no noise. Therefore, if the oracle and QAA circuit share ancilla qubits, four ancilla qubits are needed.

As the number of marked solutions increases, the number of two-qubit gates required for the oracle circuit changes. Table \ref{tab:2Qgate_each_sol} shows the minimum required number of ancilla qubits and two-qubit gates based on the number of marked solutions in the six-qubit Grover's algorithm. The oracle circuit and QAA circuit share ancilla qubits.
As the number of marked solutions increases to 1, 2, 4, 8, and 16, the minimum number of multi-controlled-gate required for the oracle circuit becomes \(C5Z\), \(C4Z\), \(C3Z\), \(C2Z\), and \(CZ\), respectively. Therefore, the number of required two-qubit gates becomes 50, 44, 38, 32, and 26, respectively.
In this paper, we conducted experiments using three \(CZ\) gates on an oracle circuit with 16 marked solutions. Therefore, we performed experiments on a six-qubit Grover's algorithm with a total of 28 two-qubit gates.
\begin{table}[h]
    \centering
    \caption{Minimum required number of ancilla qubits and two-qubit gates based on the number of marked solutions in the six-qubit Grover's algorithm.}
    \begin{tabular}{|c||c|c|c|c|c|c|}
    \hline
    \multicolumn{1}{|c||}{Number of} & \multicolumn{3}{c|}{Number of}& \multicolumn{3}{c|}{Number of}\\
    \multicolumn{1}{|c||}{marked} & \multicolumn{3}{c|}{ancilla qubits}& \multicolumn{3}{c|}{two-qubit gates}\\
    \cline{2-7}
    solutions&Oracle&QAA&Total&Oracle&QAA&Total \\ 
    \hline
     1 & 4&4&4&25&25&50\\ 
     2 & 3&4&4&19&25&44\\ 
     4 & 2&4&4&13&25&38\\ 
     8 & 1&4&4& 7&25&32\\ 
     16& 0&4&4& 1&25&26\\ 
    \hline
    \end{tabular}
    \label{tab:2Qgate_each_sol}
\end{table}

\subsection{Native gate decomposition in a trapped-ion device}
\begin{figure}
    \centering
    \includegraphics[scale=0.8]{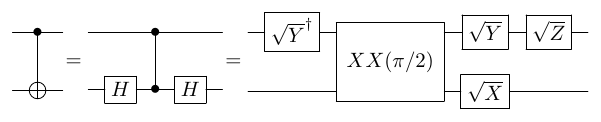}
    \caption{Decomposed quantum circuit of \(CX\) gate with \(CZ\) or \(XX(\pi/2)\) gate.}
    \label{fig:cx_cz2xx}
\end{figure}

The quantum circuit of the Grover's algorithm requires multiple single-qubit gates, such as \(H\), \(X\), \(T\)(\(\sqrt[4]{Z}\)), and \(T^{\dagger}\) gates. The native gates of the IonQ device used in the experiment are the single-qubit \(GPI\) and \(GPI2\) gates and two-qubit M{\o}lmer–S{\o}rensen (\(MS\)) gate \cite{ionq_native_gate,ionq_native_gate_journal}. The definitions of each native gate and their decompositions into the \(Z\) and \(X\) gates are shown in Eq.~\eqref{eq:GPI}, \eqref{eq:GPI2}, and \eqref{eq:MS}. The \(GPI\) and \(GPI2\) gates are the composite gates of the \(Z\) and \(X\) rotations, but the \(X\) rotation angles are limited to \(X(\pi)=X\) or \(X(\pi/2)=\sqrt{X}\). To execute quantum circuits on a trapped-ion device, all gates must be decomposed and implemented in an equivalent circuit with native gates.
\begin{align}
    \centering
    GPI({\varphi})=&
      \begin{bmatrix}
       0  &  e^{-i\varphi}\\
       e^{i\varphi} & 0\\ 
       \end{bmatrix}\nonumber\\
       =&X \cdot Z(-2{\varphi}).   \label{eq:GPI}\\
    GPI2({\varphi})=&
     \frac{1}{\sqrt{2}}
      \begin{bmatrix}
       1  &  -ie^{-i\varphi}\\
       -ie^{i\varphi} & 1\\ 
       \end{bmatrix}\nonumber\\
       =&Z({\varphi}) \cdot  \sqrt{X} \cdot Z({-\varphi}). \label{eq:GPI2}\\
    MS({\varphi}_0, {\varphi}_1)=&
     \frac{1}{\sqrt{2}}
      \begin{bmatrix}
        1  &  0  & 0 &\delta_0\\
        0  &  1  & \delta_1 & 0\\
        0  & \delta_2 & 1& 0\\
        \delta_3 & 0 &  0  & 1\\
      \end{bmatrix}, \label{eq:MS}\\
    \delta_0=&-ie^{-i({\varphi}_0+{\varphi}_1)}, \delta_1=-ie^{-i({\varphi}_0-{\varphi}_1)},\nonumber\\
    \delta_2=&-ie^{i({\varphi}_0-{\varphi}_1)}, \delta_3=-ie^{i({\varphi}_0+{\varphi}_1)}. \nonumber
\end{align}
where \(\varphi_{0}=\varphi_{1}=0\), and the \(MS\) gate is equivalent to the \(XX(\pi/2)\) gate, as shown in Eq.~\eqref{eq:MS00}.

\begin{align}
 MS(0, 0) =&
 \frac{1}{\sqrt{2}}
  \begin{bmatrix}
    1  &  0  & 0 &-i\\
    0  &  1  & -i &0\\
    0  &-i &1 &0\\
    -i& 0 &  0  & 1\\
   \end{bmatrix} \nonumber \\
   =& XX(\frac{{\pi}}{2}).
    \label{eq:MS00}
\end{align}
The procedure for implementing the quantum circuit of the Grover's algorithm on the IonQ device is described below. First, as shown in Figure~\ref{fig:cx_cz2xx}, all two-qubit gates, including the \(CX\) and \(CZ\) gates, are decomposed quantum circuits using \(XX(\pi/2)\) gate and additional single-qubit gates. Furthermore, consecutive single-qubit gates between two-qubit gates are replaced with a single-qubit gate to reduce the number of gate operations to be executed.

\begin{figure*}
    \centering
    \subfloat[Original quantum circuit.]{\includegraphics[scale=0.8]{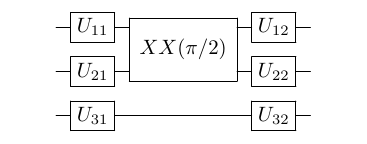}
    \label{fig:universalxx}}
    \vspace{0cm}
    \centering
    \subfloat[Quantum circuit implemented with \(ZXZXZ\) decomposition.]{\includegraphics[scale=0.8]{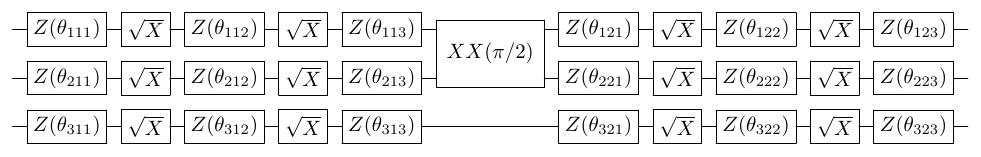}
    \label{fig:zxzxzs1}}
    \vspace{0cm}
    \centering
    \subfloat[Quantum circuit implemented with native gates up to the left side of the dotted line.]{\includegraphics[scale=0.8]{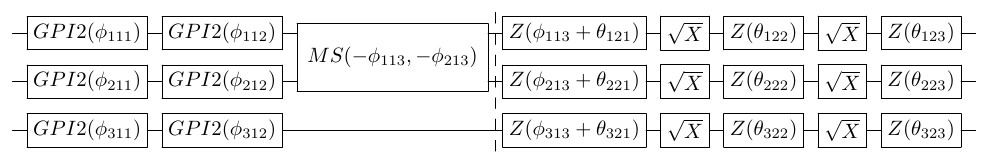}
    \label{fig:zxzxzs3}}
    \vspace{0cm}
    \centering
    \subfloat[Quantum circuits implemented with native gates for all gates.]{\includegraphics[scale=0.8]{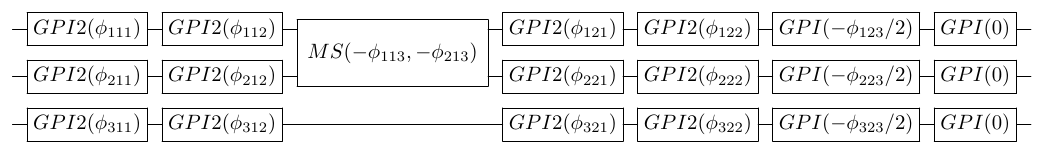}
    \label{fig:zxzxzs4}}
    \caption{Quantum circuit implementation sequence using IonQ native gates..}
    \label{fig:zxzxzsall}
\end{figure*}

For an original circuit with the universal single-qubit gates and a \(XX(\pi/2)\) gate, this decomposition is shown in Figure~\ref{fig:universalxx}, where \(U_{ij}\) is a universal single-qubit gate, where \(i\) represents the qubit number and \(j\) indicates the order of the single-qubit gate.
This figure shows a quantum circuit with and without \(XX\) gates between single-qubit gates. An arbitrary single-qubit gate can be implemented with three Euler rotation-angle \cite{OneQubitEulerDecomposer, euler_decomposition_2}. In general, at least two arbitrary rotation angles are required, but the \(GPI\) and \(GPI2\) gates allow arbitrary \(Z\) rotations; meanwhile, \(X\) rotations are limited to \(X\) and \(\sqrt{X}\). An arbitrary single-qubit gate can be implemented on a quantum device via \(ZXZXZ\) decomposition using arbitrary \(Z\) rotations and \(\sqrt{X}\) gates. As shown in Eq.~\eqref{eq:zsxzsxz}, a universal single-qubit gate \(U\) can be implemented with \(Z\) gates having three arbitrary rotation angles (\(\theta_{1}\), \(\theta_{2}\), and \(\theta_{3}\)) and two \(\sqrt{X}\) gates.
\begin{equation}
    U= Z({\theta}_{3})\cdot \sqrt{X}\cdot Z({\theta}_{2})\cdot \sqrt{X}\cdot Z({\theta}_{1}).
    \label{eq:zsxzsxz}
\end{equation}
As shown in Eq.~\eqref{eq:GPI2}, the \(GPI2\) gate includes the \(\sqrt{X}\) gate operation. Therefore, the above gate operation can be partially implemented with two serial \(GPI2\) gates as follows; however, an additional \(Z(\phi_{3})\) gate is required, as shown in Eq.~\eqref{eq:gpi2gpi2z}.
\begin{align}
    U=Z&({\phi}_{3})\cdot GPI2({\phi}_{2})\cdot GPI2({\phi}_{1})\nonumber\\
    =Z&({\phi}_{3})\cdot{\{}Z({\phi}_{2})\cdot \sqrt{X}\cdot Z(-{\phi}_{2}){\}}\nonumber\\
    &\cdot{\{}Z({\phi}_{1})\cdot \sqrt{X}\cdot Z(-{\phi}_{1}){\}}\nonumber\\
    =Z&({\phi}_{2}+{\phi}_{3})\cdot \sqrt{X}\cdot Z({\phi}_{1}-{\phi}_{2})\cdot \sqrt{X} \cdot Z(-{\phi}_{1}).
    \label{eq:gpi2gpi2z}
\end{align}
where the rotation angles \(\theta_{1}\), \(\theta_{2}\), and \(\theta_{3}\) in \(ZXZXZ\) are converted to \(\phi_{1}\), \(\phi_{2}\), and \(\phi_{3}\), respectively, for implementing with the \(GPI2\) and the residual \(Z\) gate. Where, the value of each \(\phi\) values for the single-qubit gate are Eq.~\eqref{eq:GPI2GPI2z_2}.
\begin{align}
    {\phi}_{1}&= -{\theta}_{1}=-\sum_{k=1}^{1}{\theta}_{k},\nonumber\\
    {\phi}_{2}&= -({\theta}_{1} + {\theta}_{2})=-\sum_{k=1}^{2}{\theta}_{k},\nonumber\\
    {\phi}_{3}&={\theta}_{1}+{\theta}_{2}+{\theta}_{3}=\sum_{k=1}^{3}{\theta}_{k}.
    \label{eq:GPI2GPI2z_2}
\end{align}
\begin{figure*}
    \centering
    \subfloat[Equivalent quantum circuit of \(MS\) gate using a \(XX(\pi/2)\) gate with \(Z\) gates.]{\includegraphics[scale=0.8]{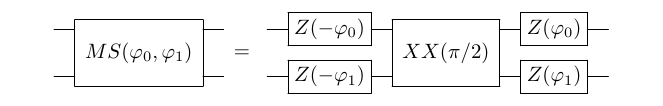}
    \label{fig:ms_gate_1}}
    \vspace{0cm}
    \subfloat[Decomposed quantum circuit reducing the number of \(Z\) gates using \(MS\) gate.]{\includegraphics[scale=0.8]{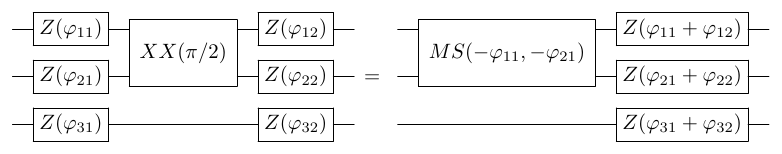}
    \label{fig:ms_gate_2}}
    \caption{M{\o}lmer–S{\o}rense (\(MS\)) gate decomposition.}
    \label{fig:ms_gate_0}
\end{figure*}
Figure~\ref{fig:zxzxzs1} shows a circuit in which universal single-qubit gates \(U_{ij}\) implemented with \(ZXZXZ\) decomposition. As shown in Figure~\ref{fig:ms_gate_1}, the \(MS(\varphi_{0}, \varphi_{1})\) gate can be decomposed to a \(XX(\pi/2)\) gate and four \(Z\) gates of the rotation angle \(\varphi_0\) and \(\varphi_1\). Thus, equivalent circuit decomposition shown in Figure~\ref{fig:ms_gate_2} can be performed. Based on Eq.~\eqref{eq:GPI2GPI2z_2} and equivalent circuit in Figure~\ref{fig:ms_gate_2}, the circuit can be decomposed into the circuit shown to the left of the dotted line in Figure~\ref{fig:zxzxzs3}, which consists of native gates (\(MS\), \(GPI2\)). However, this decomposition requires an additional rotation angle \({\phi}_{ij3}\) to be added to the first \(Z\) gate of the next cycle in \(ZXZXZ\) decomposition. These implementations are also necessary for qubits without \(XX\) gates. Thus, by propagating the residual \(Z\) rotation angle to the next \(ZXZXZ\) decomposition cycle, an arbitrary single-qubit gate between \(XX\) gates can be executed using two \(GPI2\) gates.

In this case, the \(ZXZXZ\) decomposition cycle is always inserted before measurement. Residual \(Z\) gates are always generated at the end of the last \(ZXZXZ\) decomposition cycle. To execute residual \(Z\) gates, implementation using \(GPI\) gates is considered. As shown in Eq.~\eqref{eq:redisualz_GPI}, the execution of two \(GPI\) gates at angles \(-\varphi/2\) and \(0\) equals \(Z(\varphi)\). By adding these operations, all gate operations can be implemented using the native gates of the IonQ device. Utilizing these implementing methods, the number of gate counts can be minimized.
\begin{align}
    GPI(0)\cdot GPI(-\frac{{\varphi}}{2})&={\{}X\cdot Z(0){\}}\cdot {\{}X\cdot Z({\varphi}){\}}\nonumber\\
    &=Z({\varphi}).
    \label{eq:redisualz_GPI}
\end{align}
Figure~\ref{fig:zxzxzs4} shows a circuit decomposed to only native gates (\(GPI\), \(GPI2\), and \(MS\)) finally implemented on the quantum device. The rotation angles \({\phi}_{ij1}\), \({\phi}_{ij2}\), and \({\phi}_{ij3}\) in this decomposition can be obtained using Eq.~\eqref{eq:native_gate_angle}, where \(j\) is 2 or greater.
The first term of each in Eq.~\eqref{eq:native_gate_angle} is the sum of the \(Z\) gate operation angles of the gates before the \(j\)-th single-qubit gate.
\begin{align}
    {\phi}_{ij1}&=
            -\lparen\sum_{l=1}^{j-1}\sum_{k=1}^{3}{\theta}_{ilk}+\sum_{k=1}^{1}{\theta}_{ijk}\rparen,\nonumber\\
    {\phi}_{ij2}&=
            -\lparen\sum_{l=1}^{j-1}\sum_{k=1}^{3}{\theta}_{ilk}+\sum_{k=1}^{2}{\theta}_{ijk}\rparen,\nonumber\\ 
    {\phi}_{ij3}&=\sum_{l=1}^{j-1}\sum_{k=1}^{3}{\theta}_{ilk}+\sum_{k=1}^{3}{\theta}_{ijk}=\sum_{l=1}^{j}\sum_{k=1}^{3}{\theta}_{ilk}.  \label{eq:native_gate_angle}
\end{align}

\printbibliography

@inproceedings{grover_algorithm,
  title={A fast quantum mechanical algorithm for database search},
  author={Grover, Lov K},
  booktitle={Proceedings of the Twenty-Eighth Annual ACM Symposium on Theory of Computing},
  pages={212-219},
  year={1996}
}

@article{grover_algorithm2,
  title = {Quantum Mechanics Helps in Searching for a Needle in a Haystack},
  author={Grover, Lov K.},
  journal={Physical Review Letters},
  volume = {79},
  pages = {325--328},
  year = {1997},
  publisher={American Physical Society}
}

@article{rc_qft_exp,
  title = {Randomized compiling for scalable quantum computing on a noisy superconducting quantum processor},
  author = {Hashim, Akel and Naik, Ravi K. and Morvan, Alexis and Ville, Jean-Loup and Mitchell, Bradley and Kreikebaum, John Mark and Davis, Marc and Smith, Ethan and Iancu, Costin and O'Brien, Kevin P. and Hincks, Ian and Wallman, Joel J. and Emerson, Joseph and Siddiqi, Irfan},
  journal = {Physical Review X},
  volume = {11},
  pages = {041039},
  numpages = {12},
  year = {2021},
  publisher = {American Physical Society},
}

@article{emerson_rc,
  title = {Noise tailoring for scalable quantum computation via randomized compiling},
  author = {Wallman, Joel J. and Emerson, Joseph},
  journal = {Physical Review A},
  volume = {94},
  pages = {052325},
  numpages = {9},
  year = {2016},
  publisher = {American Physical Society},
}

@article{rc_1,
  title = {Leveraging randomized compiling for the quantum imaginary-time-evolution algorithm},
  author = {Ville, Jean-Loup and Morvan, Alexis and Hashim, Akel and Naik, Ravi K. and Lu, Marie and Mitchell, Bradley and Kreikebaum, John-Mark and O'Brien, Kevin P. and Wallman, Joel J. and Hincks, Ian and Emerson, Joseph and Smith, Ethan and Younis, Ed and Iancu, Costin and Santiago, David I. and Siddiqi, Irfan},
  journal = {Physical Review Research},
  volume = {4},
  pages = {033140},
  numpages = {10},
  year = {2022},
  publisher = {American Physical Society},
}

@article{rc_pauli_frame,
  title = {Experimental {P}auli-frame randomization on a superconducting qubit},
  author = {Ware, Matthew and Ribeill, Guilhem and Rist\`e, Diego and Ryan, Colm A. and Johnson, Blake and da Silva, Marcus P.},
  journal = {Physical Review A},
  volume = {103},
  pages = {042604},
  numpages = {9},
  year = {2021},
  publisher = {American Physical Society},
}

@article{kuritaznerc,
  title = {Synergetic quantum error mitigation by randomized compiling and zero-noise extrapolation for the variational quantum eigensolver},
  author = {Kurita, Tomochika and Qassim, Hammam and Ishii, Masatoshi and Oshima, Hirotaka and Sato, Shintaro and Emerson, Joseph},
  journal = {Quantum},
  publisher = {Verein zur F{\"{o}}rderung des Open Access Publizierens in den Quantenwissenschaften},
  volume = {7},
  pages = {1184},
  year = {2023}
}

@misc{toffoli_to_cx_decomp,
      title={On the CNOT-cost of TOFFOLI gates}, 
      author={Vivek V. Shende and Igor L. Markov},
      year={2008},
      eprint={0803.2316},
      archivePrefix={arXiv},
}

@article{cnz2rptoffoli_gate,
  title = {Advantages of using relative-phase Toffoli gates with an application to multiple control Toffoli optimization},
  author = {Maslov, Dmitri},
  journal = {Physical Review A},
  volume = {93},
  pages = {022311},
  numpages = {12},
  year = {2016},
  publisher = {American Physical Society},
}

@article{OneQubitEulerDecomposer,
  title = {Improving quantum gate fidelities using optimized {E}uler angles},
  author = {Chatzisavvas, K. Ch. and Chadzitaskos, G. and Daskaloyannis, C. and Schirmer, S. G.},
  journal = {Physical Review A},
  volume = {80},
  pages = {052329},
  numpages = {11},
  year = {2009},
  publisher = {American Physical Society},
}

@article{zne_1,
  title = {Error Mitigation for Short-Depth Quantum Circuits},
  author = {Temme, Kristan and Bravyi, Sergey and Gambetta, Jay M.},
  journal = {Physical Review Letters},
  volume = {119},
  pages = {180509},
  numpages = {5},
  year = {2017},
  publisher = {American Physical Society},
}

@article{QMLEM,
  title={Machine learning for practical quantum error mitigation},
  volume={6},
  number={},
  journal={Nature Machine Intelligence},
  publisher={Springer Science and Business Media LLC},
  author={Liao, Haoran and Wang, Derek S. and Sitdikov, Iskandar and Salcedo, Ciro and Seif, Alireza and Minev, Zlatko K.},
  year={2024},
  pages={1478–1486},
}

@article{PEC,
  title = {Probabilistic error cancellation with sparse {P}auli–Lindblad models on noisy quantum processors},
  author = {van den Berg, Ewout and Minev, Zlatko K. and Kandala, Abhinav and Temme, Kristan},
  journal = {Nature Physics},
  number = {},
  volume = {19},
  pages = {1116–1121},
  numpages = {5},
  year = {2023},
}

@article{paraerrorcancel_zhang,
  title={Error-mitigated quantum gates exceeding physical fidelities in a trapped-ion system},
  author={Zhang, Shuaining and Lu, Yao and Zhang, Kuan and Chen, Wentao and Li, Ying and Zhang, Jing-Ning and Kim, Kihwan},
  journal={Nature Communications},
  volume={11},
  pages={587},
  year={2020},
  publisher={Nature Publishing Group UK London},
}

@article{grover_3qubit_exp,
  title={Complete 3-qubit {G}rover search on a programmable quantum computer},
  author={Figgatt, Caroline and Maslov, Dmitri and Landsman, Kevin A and Linke, Norbert M and Debnath, Shantanu and Monroe, Christofer},
  journal={Nature Communications},
  volume={8},
  pages={1918},
  year={2017},
  publisher={Nature Publishing Group UK London}
}

@INPROCEEDINGS{grover_ibm_exp,
  author={Mandviwalla, Aamir and Ohshiro, Keita and Ji, Bo},
  booktitle={2018 IEEE International Conference on Big Data (Big Data)}, 
  title={Implementing {G}rover’s algorithm on the {IBM} quantum computers}, 
  year={2018},
  volume={},
  number={},
  pages={2531-2537},
}

@article{grover_exp,
  title = {Implementation of {G}rover's quantum search algorithm in a scalable system},
  author = {Brickman, K.-A. and Haljan, P. C. and Lee, P. J. and Acton, M. and Deslauriers, L. and Monroe, C.},
  journal = {Physical Review A},
  volume = {72},
  pages = {050306},
  numpages = {4},
  year = {2005},
  publisher = {American Physical Society},
}

@article{post_selection,
author = {Endo ,Suguru and Cai ,Zhenyu and Benjamin ,Simon C. and Yuan ,Xiao},
title = {Hybrid Quantum-Classical Algorithms and Quantum Error Mitigation},
journal = {Journal of the Physical Society of Japan},
volume = {90},
number = {3},
pages = {032001},
year = {2021},
}

@misc{Knill,
      title={Quantum Computing with Very Noisy Devices},
      author={Knill, E.},
      year={2004},
      eprint={quant-ph/0410199},
      archivePrefix={arXiv},
}

@article{Kern,
   title={Quantum error correction of coherent errors by randomization},
   volume={32},
   number={1},
   journal={The European Physical Journal D},
   publisher={Springer Science and Business Media LLC},
   author={Kern, O. and Alber, G. and Shepelyansky, D. L.},
   year={2005},
   pages={153–156},
}

@online{ionq_native_gate,
    title = {Native Gates},
    year      = {2026},
    url = {https://docs.ionq.com/guides/getting-started-with-native-gates},
}

@online{ionq_aria,
    title = {Ion{Q} {A}ria},
    year      = {2026},
    url = {https://ionq.com/quantum-systems/aria},
}

@ARTICLE{ionq_native_gate_journal,
  author={Saki, Abdullah Ash and Topaloglu, Rasit Onur and Ghosh, Swaroop},
  journal={IEEE Access},
  title={Shuttle-Exploiting Attacks and Their Defenses in Trapped-Ion Quantum Computers},
  year={2022},
  volume={10},
  number={},
  pages={2686-2699},
}

@article{CnXgate_decomposition_with_ancilla,
  title = {Elementary gates for quantum computation},
  author = {Barenco, Adriano and Bennett, Charles H. and Cleve, Richard and DiVincenzo, David P. and Margolus, Norman and Shor, Peter and Sleator, Tycho and Smolin, John A. and Weinfurter, Harald},
  journal = {Physical Review A},
  volume = {52},
  pages = {3457--3467},
  numpages = {0},
  year = {1995},
  publisher = {American Physical Society},
}

@article{CnZgate_decomposition_without_ancilla,
  title = {Decompositions of multiple controlled-$Z$ gates on various qubit-coupling graphs},
  author = {Nakanishi, Ken M. and Satoh, Takahiko and Todo, Synge},
  journal = {Physical Review A},
  volume = {110},
  pages = {012604},
  numpages = {7},
  year = {2024},
  publisher = {American Physical Society},
}

@article{euler_decomposition_2,
  title = {Error rate reduction of single-qubit gates via noise-aware decomposition into native gates},
  author = {Maldonado, Thomas J and Flick, Johannes and Krastanov, Stefan and Galda, Alexey},
  journal = {Scientific Reports},
  volume = {12},
  pages = {6379},
  year = {2022},
}

@article{qaa_ref0,
  title = {Arbitrary phases in quantum amplitude amplification},
  author = {H\o{}yer, Peter},
  journal = {Physical Review A},
  volume = {62},
  pages = {052304},
  numpages = {5},
  year = {2000},
  publisher = {American Physical Society},
}

@article{qaa_ref1,
  title = {Gravitational wave matched filtering by quantum {M}onte {C}arlo integration and quantum amplitude amplification},
  author = {Miyamoto, Koichi and Morr\'as, Gonzalo and Yamamoto, Takahiro S. and Kuroyanagi, Sachiko and Nesseris, Savvas},
  journal = {Physical Review Research},
  volume = {4},
  pages = {033150},
  numpages = {20},
  year = {2022},
  publisher = {American Physical Society},
}

@article{qaa_ref2,
  title = {Quantum multiplication algorithm based on the convolution theorem},
  author = {Ramezani, Mehdi and Nikaeen, Morteza and Farman, Farnaz and Ashrafi, Seyed Mahmoud and Bahrampour, Alireza},
  journal = {Physical Review A},
  volume = {108},
  pages = {052405},
  numpages = {10},
  year = {2023},
  publisher = {American Physical Society},
}

@article{qaa_ref3,
  title = {Amplitude estimation without phase estimation},
  author = {Suzuki, Yohichi and Uno, Shumpei and Raymond, Rudy and Tanaka, Tomoki and Onodera, Tamiya and Yamamoto, Naoki},
  journal = {Quantum Information Processing},
  volume = {19},
  pages = {75},
  year = {2020},
  publisher = {Springer},
}

@article{qaa_ref4,
  title = {Quadratic acceleration of multistep probabilistic algorithms for state preparation},
  author = {Nishi, Hirofumi and Kosugi, Taichi and Nishiya, Yusuke and Matsushita, Yuichiro},
  journal = {Physical Review Research},
  volume = {6},
  pages = {L022041},
  numpages = {6},
  year = {2024},
  publisher = {American Physical Society},
}

@article{post_processing_protocol,
  title = {Low-cost error mitigation by symmetry verification},
  author = {Bonet-Monroig, X. and Sagastizabal, R. and Singh, M. and O'Brien, T. E.},
  journal = {Physical Review A},
  volume = {98},
  pages = {062339},
  numpages = {10},
  year = {2018},
  publisher = {American Physical Society},
}

@misc{grover_oracle_logic,
      author = {Peter Schwabe and Bas Westerbaan},
      title = {Solving binary {MQ} with Grover's algorithm},
      howpublished = {Cryptology {ePrint} Archive, Paper 2019/151},
      year = {2019},
}

@article{kim_pauli_twirling,
  title = {Scalable error mitigation for noisy quantum circuits produces competitive expectation values},
  author = {Kim, Youngseok and Wood, Christopher J. and Yoder, Theodore J. and Merkel, Seth T. and Gambetta, Jay M. and Temme, Kristan and Kandala, Abhinav},
  journal = {Nature Physics},
  volume = {19},
  pages = {752--759},
  year = {2023},
}
\end{document}